# Continuous Time Markov Chains for Analysis of Non-Alcoholic Fatty liver Disease Evolution

Iman M. Attia *

Imanattiathesis1972@gmail.com , imanattia1972@gmail.com

*Department of Mathematical Statistics , Faculty of Graduate Studies for Statistical Research , Cairo University , Egypt

*Abstract-*In the present paper, progression of non-alcoholic fatty liver disease (NAFLD) process is modeled by Continuous time Markov chains (CTMC) with 4 states .The transition intensities among the states are estimated using maximum likelihood estimation (MLE) method. The transition probabilities are also calculated. The mean sojourn time and its variance are estimated as well as the state probability distribution and its asymptotic covariance matrix. The life expectancy of the patient, one of the important statistical indices, is also obtained. The paper illustrates the new approach of using MLE to compensate for missing values in the follow up periods of patients in the longitudinal studies. This new approach also yields that the estimated rates among states are approximately equals to the observed rates.

## I. INTRODUCTION

CTMC is frequently used to model panel data in various fields of science, including: medicine, sociology, biology, physics and finance. It is one of the most common used tools to model disease progression and evolution over time periods. In medical research studies, this technique is used to model illness-death process in which each patient starts in one initial state and eventually ends in absorbing or final state .It has been addressed by many authors in the medical field such as: Estes et al.[1] used multistate Markov chains to model the epidemic of nonalcoholic fatty liver disease. Younossi et al. [2] used the multistate Markov chains to demonstrate the economic and clinical burden of nonalcoholic fatty liver disease in United States and Europe. Anwar & Mahmoud [3] used CTMC to model chronic renal failure in patients. Grover et al. [4] used time dependent multistate Markov chains to assess progression of liver cirrhosis in patients with various prognostic factors. Bartolomeo et al. [5] employed a hidden Markov model to study progression of liver cirrhosis to hepatocellular carcinoma and death. Saint-Pierre et al. [6] used CTMC to study asthma disease process with time dependent covariates. Klotz & Sharples [7] modeled the follow up of patients with heart transplants using multistate Markov chains.

Studying natural history of disease during which individuals start at one initial state then as time passes the patients move from one state to another, can be investigated by using multistate Markov chains. Evolution of the disease over different phases can be monitored by taking repeated observations of the disease stage at pre-specified time points following entry into the study. Disease stage is recorded at time of observation while the exact time of state change is unobserved. NAFLD is a multistage disease process; in its simplest form has a general structure model as depicted in Figure 1.

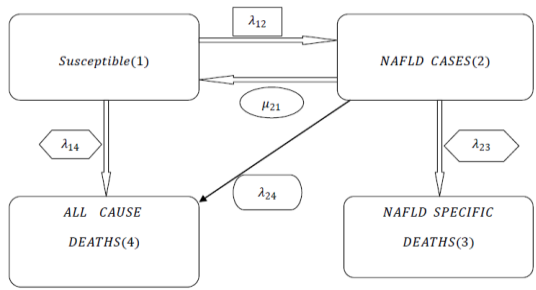

Figure 1 : General Model Structure

NAFLD stages are modeled as time homogenous CTMC , that is to mean $P_{ij}(\Delta t)$ depends on $\Delta t$ and not on $t$ ,with constant transition intensities $\lambda_{ij}$ over time, exponentially distributed time spent within each state and patients' events follow Poisson distribution. The states are: one for the susceptible cases (state 1) and one for NAFLD cases (state 2) and two absorbing states ; one for the death due to NAFLD (state 3) and one for death due to any other cause (state 4). The transition rate $\lambda_{12}$ is the rate of progression from state 1 to state 2, while the transition rate $\mu_{12}$ is the regression rate from state 2 to state 1. The transition rate $\lambda_{23}$ is the progression rate from state 2 to state 3 and $\lambda_{24}$ is the rate of progression from state 2 to state 4.

For simplicity, all individuals are assumed to enter the disease process at stage one and they are all followed up with the same length of time interval between measurements. According to American Association for Study of Liver Disease , American College of Gastroenterology, and the American Gastroenterological Association, NAFLD to be defined requires (a) there is evidence of hepatic steatosis (HS) either by imaging or by histology and (b)there are no causes for secondary hepatic fat accumulation such as significant alcohol consumption, use of steatogenic medications or hereditary disorders [8].This is the same definition established by European Association for the Study of the Liver (EASL),European Association for the Study of Diabetes (EASD)and European Association for the Study of Obesity(EASO)[9]. NAFLD can be categorized histologically into nonalcoholic fatty liver (NAFL) or nonalcoholic steato-hepatitis (NASH). NALF is defined as the presence of ≥ 5% (HS) without evidence of hepatocellular injury in the form of hepatocyte ballooning .NASH is defined as the presence of ≥ 5 % HS and inflammation with hepatocyte injury (ballooning), with or without any fibrosis. Liver biopsy is presently the most trustworthy procedure for diagnosing the presence of steatohepatitis (HS) and fibrosis in NAFLD patients [10]. The limitations of this procedure are cost, sampling error, and procedure-related morbidity and mortality. MR imaging, by spectroscopy[11] or by proton density fat fraction[12], is an excellent noninvasive technique for quantifying HS and is being widely used in NAFLD clinical trials [13]. The use of transient elastrography (TE) to obtain continuous attenuation parameters is a promising tool for quantifying hepatic fat in an ambulatory setting [14]. However, quantifying noninvasively HS in patients with NAFLD is limited in routine clinical care. The susceptible cases have risk factors for developing NAFLD such as visceral obesity, type 2 diabetes mellitus (T2DM), dyslipidemia, older age, male sex and being of Hispanic ethnicity [15].

The paper is divided into 7 sections. In section I the transition probabilities and transition rates are thoroughly discussed. In section II mean sojourn time and its variance are reviewed. In section III state probability distribution and its covariance matrix are discussed. While in section IV the life expectancy of the patients are considered. In section V expected numbers of patients in each state is obtain. A hypothetical numerical example is used in section VI to illustrate the above concepts. Lastly a brief summary is comprehended in section VII.

I. Transition Rates and Probabilities

NAFLD is modeled by a multistate Markov chains which define a stochastic process
$[(X(t), t \in T)]$ over a finite state space $S$
$= \{1,2,3,4\}$ and $T = [0,t]$ and $t < \infty$

The transitions can occur at any point in time and hence called continuous time Markov chains in contrast to the discrete time Markov chains in which transitions occur at fixed points in time. The rates at which these transitions occur are constant over time and thus are independent of t that is to say the transition of patient from $state\ i\ at\ time = t\ to\ state\ j\ at\ t = t + s\ where\ s = \Delta t$ depends on difference between two consecutive time points. And it's defined as $\theta_{ij}(t) = \lim_{\Delta t \to 0} \frac{P_{ij}(\Delta t) - I}{\Delta t}$ or the Q matrix.

For the above multistate Markov model demonstrating the NAFLD disease process; the forward Kolomogrov differential equations are the following:

$\frac{d}{dt}P_{ij}(t)$
$= \begin{bmatrix} P_{11} & P_{12} & P_{13} & P_{14} \\ P_{21} & P_{22} & P_{23} & P_{24} \\ 0 & 0 & P_{33} & 0 \\ 0 & 0 & 0 & P_{44} \end{bmatrix} \begin{bmatrix} -(\lambda_{12} + \lambda_{14}) & \lambda_{12} & 0 & \lambda_{14} \\ \mu_{21} & -(\mu_{21} + \lambda_{23} + \lambda_{24}) & \lambda_{23} & \lambda_{24} \\ 0 & 0 & 0 & 0 \\ 0 & 0 & 0 & 0 \end{bmatrix}$

The Kolmogrove differential equations:

$\frac{dP_{11}}{dt} = -P_{11}(\lambda_{12} + \lambda_{14}) + P_{12}\mu_{21}$

$\frac{dP_{12}}{dt} = P_{11}\lambda_{12} - P_{12}(\mu_{21} + \lambda_{23} + \lambda_{24})$

$\frac{dP_{13}}{dt} = P_{12}\lambda_{23}$

$\frac{dP_{14}}{dt} = P_{11}\lambda_{14} + P_{12}\lambda_{24}$

$\frac{dP_{21}}{dt} = -P_{21}(\lambda_{12} + \lambda_{14}) + P_{22}\mu_{21}$

$\frac{dP_{22}}{dt} = P_{21}\lambda_{12} - P_{22}(\mu_{21} + \lambda_{23} + \lambda_{24})$

$\frac{dP_{23}}{dt} = P_{22}\lambda_{23}$

$\frac{dP_{24}}{dt} = P_{21}\lambda_{14} + P_{22}\lambda_{24}$

$P_{33} = 1$
$P_{44} = 1$

The solution of this system of equations will give the $P_{ij}(t)$ (see appendix section I)

$P_{ij}(t) = \begin{bmatrix} P_{11} & P_{12} & P_{13} & P_{14} \\ P_{21} & P_{22} & P_{23} & P_{24} \\ 0 & 0 & P_{33} & 0 \\ 0 & 0 & 0 & P_{44} \end{bmatrix}$

$P_{ij}(t)$ satisfies the following properties:
1. $P_{ij}(t+s) = \sum_{i,j,l \in S} P_{il}(t)P_{lj}(s)$, $\forall t \geq 0, s \geq 0, i,j,l \in S$; obying kolmogrov equations
2. $\sum_S P_{ij}(t) = 1$
3. $P_{ij}(t) \geq 0$, $\forall t \geq 0$ and $i,j \in S$

While the Q matrix satisfies the following conditions:
1. $\sum_S q_{ij}(t) = 0$
2. $q_{ij}(t) \geq 0$, $i \neq j$
3. $-\sum_S q_{ij}(t) = q_{ii}$, $i = j$

Where the $q_{ij}$ is the $(i,j)$ th entry in the Q matrix emphasizing that the $P_{ij}$ depends only on the interval between $t_1$ and $t_2$ not on $t_1$.

A. Maximum Likelihood Estimation of the Q Matrix

Let $n_{ijr}$ be the number of individuals in state $i$ at $t_{r-1}$ and in state $j$ at time $t_r$. Conditioning on the distribution of individuals among states at $t_0$, then the likelihood function for $\theta$ is

$L(\theta) = \prod_{r=1}^{w} \left\{ \prod_{i,j=1}^{k} [P_{ij}(t_{r-1}, t_r)]^{n_{ijr}} \right\}$,

where $k$ is the index of the number of states

$\log L(\theta) = \sum_{r=1}^{\tau} \sum_{i,j=1}^{k} n_{ijr} \log P_{ij}(t_{r-1}, t_r)$,

where $\tau = (t_r - t_{r-1})$

According to Kalbfleisch & Lawless [16], applying Quasi-Newton method to estimate the rates mandates calculating the score function which is a vector-valued function for the required rates and it's the first derivative of the probability transition function with respect to $\theta$. The second derivative is assumed to be zero.

$S(\theta) = \frac{\partial}{\partial \theta_h} \log L(\theta) = \sum_{r=1}^{\tau} \sum_{i,j=1}^{k} n_{ijr} \frac{\partial P_{ij}(\tau)/\partial \theta_h}{P_{ij}(\tau)}$,

$h = 1,2,3,4,5$ while $P_{ij}(\tau) = \frac{n_{ijr}}{n_{i+}}$

where: $\theta_1 = \lambda_{12}, \theta_2 = \lambda_{14}, \theta_3 = \mu_{21}, \theta_4 = \lambda_{23}, \theta_5 = \lambda_{24}$

$$\frac{n_{ijr}}{P_{ij}(\tau)} = n_{i+}, \quad \text{such that} \quad n_{i+} = \sum_{j=1}^{k} n_{ijr}$$

$S(\theta) = \tau e^{\Lambda\tau} d\Lambda$

*and it's scaled 4 times by $n_{1+}$ and another 4 times by $n_{2+}$ for each $\tau$*
$\Lambda$ *is the eigenvalues for each Q matrix in each $\tau$ ( see appendix Section 1 & excel sheet )*

$$\frac{\partial^2}{\partial\theta_g \partial\theta_h} \log L(\theta)$$
$$= \sum_{r=1}^{\tau} \sum_{i,j=1}^{k} n_{ijr} \left\{ \frac{\partial^2 P_{ij}(\tau)/\partial\theta_g \partial\theta_h}{P_{ij}(\tau)} - \frac{\partial P_{ij}(\tau)/\partial\theta_g \partial P_{ij}(\tau)/\partial\theta_h}{P_{ij}^2(\tau)} \right\}$$

Assuming the second derivative is zero and $\frac{n_{ijr}}{P_{ij}(\tau)} = n_{i+}$ then

$$M_{ij}(\theta) = \frac{\partial^2}{\partial\theta_g \partial\theta_h} \log L(\theta)$$
$$= -\sum_{r=1}^{\tau} \sum_{i,j=1}^{k} n_{i+} \frac{\partial P_{ij}(\tau)/\partial\theta_g \partial P_{ij}(\tau)/\partial\theta_h}{P_{ij}(\tau)}$$

The Quasi-Newton formula is
$\theta_1 = \theta_0 + [M(\theta_0)]^{-1} S(\theta_0)$

According to Klotz & Sharples [7] the initial $\theta_0 = \frac{n_{ijr}}{n_{i+}}$ for $\Delta t = 1$

According to Jackson [17] initial value for a model can be set by supposing that transitions between states take place only at the observation times. If $n_{ij}$ transitions are observed from $state\ i$ to $states\ j$ and a total of $n_i$ transitions from $state\ i$, then $\frac{q_{ij}}{q_{ii}}$ can be estimated by $\frac{n_{ij}}{n_{ii}}$. Then, given a total of $T_i$ years spent in $state\ i$, the mean sojourn time $\frac{1}{q_{ii}}$ can be estimated as $\frac{T_i}{n_i}$. thus, $\frac{n_{ij}}{T_i}$ is a crude estimate of $q_{ij}$. The Quasi-Newton method produces $\hat{\theta}$ upon convergence and $[M(\theta_0)]^{-1}$ is estimate of the asymptotic covariance matrix of $\hat{\theta}$.

For this NAFLD process  (*see appendix section I*)

## II. Mean Sojourn Time

It is the mean time spent by a patient in a given state i of the process. It is calculated in relations to transition rates $\hat{\theta}$. These times are independent and exponentially distributed random variables with mean $\frac{1}{\lambda_i}$ where $\lambda_i = -\lambda_{ii}$; $i = 1,2,3,4$. Denoting mean sojourn time by $s_i$ for state i at visits 1,2,…

$$s_1 = \frac{1}{(\lambda_{12} + \lambda_{14})}, \quad \text{and} \quad s_2 = \frac{1}{(\mu_{21} + \lambda_{23} + \lambda_{24})}$$

According to Kalbfleisch & Lawless [16] the asymptotic variance of this time is calculated by applying multivariate delta method:

$$var(s_i) = \left[\left(q_{ii}(\hat{\theta})\right)^{-2}\right]^2 \sum_{h=1}^{5} \sum_{g=1}^{5} \frac{\partial q_{ii}}{\partial \theta_g} \frac{\partial q_{ii}}{\partial \theta_h} [M(\theta)]^{-1}|_{\theta = \hat{\theta}}$$

For this NAFLD process  (*see appendix section II*)

## III. State Probability Distribution

According to Cassandras & Lafortune [18] it is the probability distribution for each state at a specific time point given the initial probability distribution. Thus using the rule of total probability; a solution describing the transient behavior of a chain characterized by Q and an initial condition $\pi(0)$ is obtained by direct substitution to solve:
$\pi(t) = \pi(0)P(t)$
For this NAFLD process  (*see appendix section III*)

To obtain stationary probability distribution when $t$ goes to infinity or in other words when the process does not depend on time
$let's\ call \quad \lambda_{12} + \lambda_{14} = \gamma_1,\ \mu_{21} + \lambda_{23} + \lambda_{24} = \gamma_2$
$\pi(t) = \pi(0)P(t) = \pi(0)e^{Qt}$,
$\qquad differentiating\ both\ sides$

$$\frac{d}{dt}\pi(t)\Big|_{t=0} = \pi(0)Q$$

$$\frac{d}{dt}\pi(t)\Big|_{t=0} = [\pi_{0(1)}\ \pi_{0(2)}\ \pi_{0(3)}\ \pi_{0(4)}] \begin{bmatrix} -\gamma_1 & \lambda_{12} & 0 & \lambda_{14} \\ \mu_{21} & -\gamma_2 & \lambda_{23} & \lambda_{24} \\ 0 & 0 & 0 & 0 \\ 0 & 0 & 0 & 0 \end{bmatrix}$$

$$\frac{d}{dt}\pi_1(t)\Big|_{t=0} = -\pi_{0(1)}\gamma_1 + \pi_{0(2)}\mu_{21}$$
$$\frac{d}{dt}\pi_2(t)\Big|_{t=0} = \pi_{0(1)}\lambda_{12} - \pi_{0(2)}\gamma_2$$
$$\frac{d}{dt}\pi_3(t)\Big|_{t=0} = \pi_{0(2)}\lambda_{23}$$
$$\frac{d}{dt}\pi_4(t)\Big|_{t=0} = \pi_{0(1)}\lambda_{14} + \pi_{0(2)}\lambda_{24}$$

$By\ solving\ these\ equations, the\ vector\ [\pi_1\ \pi_2\ \pi_3\ \pi_4]$
$point\ is\ obtained\ at\ specific\ time$
Solving these differential equations even for simple chains is not a trivial matter.
$$\pi_z = \lim_{t \to \infty} \pi_z(t)$$
If this limit exists so there is a stationary or steady state distribution and as $t \to \infty$, the $\frac{d}{dt}\pi_j(t) = 0$, since $\pi_z(t)$ does not depend on time
$$\frac{d}{dt}\pi(t) = \pi(t)Q \quad will\ reduce\ to\ \pi(t)Q = 0$$
$By\ solving\ \pi Q = 0,\ subject\ to\ \sum_{all\ z} \pi_z = 1$,
$the\ state\ probability\ distribution\ is\ obtained$

$$[\pi_{(1)}\ \pi_{(2)}\ \pi_{(3)}\ \pi_{(4)}] \begin{bmatrix} -\gamma_1 & \lambda_{12} & 0 & \lambda_{14} \\ \mu_{21} & -\gamma_2 & \lambda_{23} & \lambda_{24} \\ 0 & 0 & 0 & 0 \\ 0 & 0 & 0 & 0 \end{bmatrix} = \begin{bmatrix} 0 \\ 0 \\ 0 \\ 0 \end{bmatrix}$$

$-\pi_{(1)}\gamma_1 + \pi_{(2)}\mu_{21} = 0$, $\quad \pi_{(1)}\lambda_{12} - \pi_{(2)}\gamma_2 = 0$,
$\pi_{(2)}\lambda_{23} = 0$, $\quad \pi_{(1)}\lambda_{14} + \pi_{(2)}\lambda_{24} = 0$
$1 = \pi_{(1)} + \pi_{(2)} + \pi_{(3)} + \pi_{(4)}$
The above equations are expressed in matrix notation as:

$$\begin{bmatrix} -\gamma_1 & \mu_{21} & 0 & 0 \\ \lambda_{12} & -\gamma_2 & 0 & 0 \\ 0 & \lambda_{23} & 0 & 0 \\ \lambda_{14} & \lambda_{24} & 0 & 0 \\ 1 & 1 & 1 & 1 \end{bmatrix} \begin{bmatrix} \pi_1 \\ \pi_2 \\ \pi_3 \\ \pi_4 \end{bmatrix} = \begin{bmatrix} 0 \\ 0 \\ 0 \\ 0 \\ 1 \end{bmatrix}$$

$$\begin{bmatrix} \pi_1 \\ \pi_2 \\ \pi_3 \\ \pi_4 \end{bmatrix} = \begin{bmatrix} 0 \\ 0 \\ \pi_3 \\ 1-\pi_3 \end{bmatrix}$$

A. Asymptotic Covariance of the State Probability Distribution

To obtain this, multivariate delta method is used as well as the following function of the
$Q'\pi = 0$, as $\pi$ is not a simple function of $\theta$
$$\frac{\partial}{\partial \theta_h} F(\theta_h, \pi_i) = \frac{\partial}{\partial \theta_h} Q'\pi_i = 0 \quad ,$$

with implicit differentiation ,

$$\frac{\partial}{\partial \theta_h} F(\theta_h, \pi_i) = \frac{\partial}{\partial \theta_h} Q'\pi_i = [Q'] \left[\frac{\partial}{\partial \theta_h} \pi_i\right] + \pi_i \left[\frac{\partial}{\partial \theta_h} Q'\right]^T ,$$

let's call $\pi_i \left[\frac{\partial}{\partial \theta_h} Q'\right]^T = C(\theta)$

solving for $\left[\frac{\partial}{\partial \theta_h} \pi_i\right]$

$$\left[\frac{\partial}{\partial \theta_h} \pi_i\right] = -[Q']^{-1} C(\theta)$$

Let $\left[\frac{\partial}{\partial \theta_h} \pi_i\right] = A(\theta)$

By multivariate delta method
$var(\pi) = A(\theta) var(\theta) A(\theta)'$ ,
where $var(\theta) = [M(\theta)]^{-1}$

For this NAFLD process: (see appendix section III)

IV. Life Expectancy of Patient in NAFLD Disease Process

The disease process is composed of state 1 and state 2 which are transient states, while state 3 and state 4 both are absorbing states. So partitioning the Q matrix into 4 sets

$$Q = \begin{bmatrix} -(\lambda_{12}+\lambda_{14}) & \lambda_{12} & 0 & \lambda_{14} \\ \mu_{21} & -(\mu_{21}+\lambda_{23}+\lambda_{24}) & \lambda_{23} & \lambda_{24} \\ 0 & 0 & 0 & 0 \\ 0 & 0 & 0 & 0 \end{bmatrix} = \begin{bmatrix} B & A \\ 0 & 0 \end{bmatrix}$$

Where $B = \begin{bmatrix} -(\lambda_{12}+\lambda_{14}) & \lambda_{12} \\ \mu_{21} & -(\mu_{21}+\lambda_{23}+\lambda_{24}) \end{bmatrix}$,

$A = \begin{bmatrix} 0 & \lambda_{14} \\ \lambda_{23} & \lambda_{24} \end{bmatrix}$

$A = BZ$
let $(\lambda_{12}+\lambda_{14}) = \gamma_1$, $(\mu_{21}+\lambda_{23}+\lambda_{24}) = \gamma_2$
so Z

$$= \begin{bmatrix} \frac{\lambda_{12}\lambda_{23}}{\mu_{21}\lambda_{12}-\gamma_1\gamma_2} & \frac{\lambda_{12}(\mu_{21}\lambda_{14}+\lambda_{24}\gamma_1)-\lambda_{14}(\mu_{21}\lambda_{12}-\gamma_1\gamma_2)}{\gamma_1(\mu_{21}\lambda_{12}-\gamma_1\gamma_2)} \\ \frac{\gamma_1\lambda_{23}}{\mu_{21}\lambda_{12}-\gamma_1\gamma_2} & \frac{(\mu_{21}\lambda_{14}+\lambda_{24}\gamma_1)}{\mu_{21}\lambda_{12}-\gamma_1\gamma_2} \end{bmatrix}$$

$[\dot{P}(t) \quad \dot{P}_k(t)] = [P(t) \quad P_k(t)] \begin{bmatrix} B & A \\ 0 & 0 \end{bmatrix}$ can be written as
$\dot{P}(t) = P(t)B$
$\dot{P}_k(t) = P(t)A$
The solution to $\dot{P}(t) = P(t)B$ is $P(t) = P(0)e^{BT}$
then $\dot{P}_k(t) = P(0)e^{BT}A$

$$P_k(t) = A\frac{e^{Bt}}{B}\bigg|_{t=0}^{t=t} = A\left[\frac{e^{Bt}}{B} - \frac{1}{B}\right] = \frac{A}{B}[e^{Bt}-1] = \frac{BZ}{B}[e^{Bt}-1]$$
$$= Z[e^{Bt}-1]$$

and $e^{BT} = 1 + Bt + \frac{(Bt)^2}{2!} + \frac{(Bt)^3}{3!} + \frac{(Bt)^4}{4!} + \cdots = \sum_{j=0}^{\infty} \frac{(Bt)^j}{j!}$

If $\tau_k$ is the time taken from state i to reach the absorbing death state from the initil time
$$F_k(t) = pr[\tau_k \le t] = pr[X(t) = k] = P_k(t)$$
$$= Z[e^{Bt}-1]$$

The moment theory for Laplace transform can be used to obtain the mean of the time which has the above cumulative distribution function.
CTMC can be written in a Laplace transform such that:

$[sP^*(s) - P(0) \quad sP^*_k(s)] = [P^*(s) \quad P^*_k(s)] \begin{bmatrix} B & A \\ 0 & 0 \end{bmatrix}$

$\therefore sP^*(s) - P(0) = P^*(s)B$
$\therefore sP^*_k(s) = P^*(s)A$
Rearrange :
$\therefore sP^*(s) - P^*(s)B = P(0)$
$P^*(s) [sI-B] = P(0) \rightarrow P^*(s) = P(0)[sI-B]^{-1}$
$\therefore sP^*_k(s) = P^*(s)A \rightarrow P^*_k(s) = \frac{1}{s} P^*(s)A$
$$= \frac{1}{s} P(0)[sI-B]^{-1}A$$

$F^*_k(s) = \frac{1}{s} P(0)[sI-B]^{-1}A$
$f^*_k(s) = sF^*_k(s) = P(0)[sI-B]^{-1}A$ ;
where $A = BZ$
Mean time to absorption:
$$E(\tau_k) = (-1)\frac{df^*_k(s)}{ds}\bigg|_{s=0} = P(0)[sI-B]^{-2}A|_{s=0}$$
$$= P(0)[B]^{-1}Z = [B]^{-1}Z$$
$E(\tau_k) = [B]^{-1}Z$

For this NAFLD process: (see appendix section IV )

$E(\tau_{13}) = \frac{\lambda_{12}\lambda_{23}(\gamma_1+\gamma_2)}{(\gamma_1\gamma_2-\mu_{21}\lambda_{12})^2}$

$E(\tau_{14}) = \frac{\lambda_{12}(\mu_{21}\lambda_{14}+\lambda_{24}\gamma_1)(\gamma_1+\gamma_2)}{\gamma_1 (\gamma_1\gamma_2-\mu_{21}\lambda_{12})^2} + \frac{\lambda_{14}\gamma_2}{\gamma_1(\gamma_1\gamma_2-\mu_{21}\lambda_{12})}$

$E(\tau_{23}) = \frac{\lambda_{23}(\mu_{21}\lambda_{12}+\gamma_1^2)}{(\gamma_1\gamma_2-\mu_{21}\lambda_{12})^2}$

$E(\tau_{24}) = \frac{(\mu_{21}\lambda_{14}+\lambda_{24}\gamma_1)(\mu_{21}\lambda_{12}+\gamma_1^2)}{\gamma_1 (\gamma_1\gamma_2-\mu_{21}\lambda_{12})^2} + \frac{\mu_{21}\lambda_{14}}{\gamma_1(\gamma_1\gamma_2-\mu_{21}\lambda_{12})}$

V. Expected Number of Patients in Each State

Let $u(0)$ be the size of patients in a specific state at specific time $t = 0$. The initial size of patients $u(0) = \sum_{j=1}^{2} u_j(0)$, as there are 2 transient states and 2 absorbing states, where $u_j(0)$ is the initial size or number of patients in state $j$ at time $t = 0$ given that $u_3(0) = 0$ and $u_4(0) = 0$ i.e initial size of patients in state 3 and state 4 ( both are absorbing death state) are zero at initial time point $t = 0$. As the transition or the movement of the patients among states are independent so at the end of the whole time interval $(0, t)$ and according to Chiang[19], there will be $u_j(t)$ patients in state 1 and in state 2 at time $t$ , also there will be $u_3(t)$ patients in state 3 (death state) at time $t$ and $u_4(t)$ patients in state 4 (death state) at time $t$ .

$$E[u_j(t)|u_j(0)] = \sum_{j=1, i=1}^{2} u_i(0)P_{ij}(t) , \quad i\& j = 1,2$$

$$E[u_j(t)|u_i(0)] = \sum_{j=3,i=1}^{j=4,i=2} u_i(0) P_{ij}(t),$$
$$i = 1,2 \text{ and } j = 3,4$$

In matrix notation :
$$E[u_j(t)|u_i(0)]$$
$$= [u_1(0) \quad u_2(0) \quad 0 \quad 0] \begin{bmatrix} P_{11} & P_{12} & P_{13} & P_{14} \\ P_{21} & P_{22} & P_{23} & P_{24} \\ 0 & 0 & P_{33} & 0 \\ 0 & 0 & 0 & P_{44} \end{bmatrix}$$
$$= [u_1(t) \quad u_2(t) \quad u_3(t) \quad u_4(t)]$$
$$u_1(t) = u_1(0)P_{11} + u_2(0)P_{21}$$
$$u_2(t) = u_1(0)P_{12} + u_2(0)P_{22}$$
$$u_3(t) = u_1(0)P_{13} + u_2(0)P_{23}$$
$$u_4(t) = u_1(0)P_{14} + u_2(0)P_{24}$$

## VI. Hypothetical Numerical Example

To illustrate the above concepts and discussion, a hypothetical numerical example is introduced. It does not represent real data but it is for demonstrative purposes.*( see suppl. Info. excel file )*

A study was conducted over 8 years on 310 patients with risk factors for developing NAFLD such as type 2 diabetes mellitus, obesity, and hypertension acting alone or together as a metabolic syndrome. The patients were decided to be followed up every year by a liver biopsy to identify the NAFLD cases, but the actual observations were recorded as shown in the excel sheet 1 (see supplementary material).
The estimated transition rate matrix Q is:
$$Q = \begin{bmatrix} -.31365 & .2908 & 0 & .02285 \\ .02805 & -.30365 & .2076 & .068 \\ 0 & 0 & 0 & 0 \\ 0 & 0 & 0 & 0 \end{bmatrix}$$
$$var(\theta) = \begin{bmatrix} .061475 & -.04645 & -.01585 & 0 & 0 \\ -.04645 & .037836 & -.00613 & 0 & 0 \\ -.01585 & -.00613 & .123658 & 0 & 0 \\ 0 & 0 & 0 & 0 & 0 \\ 0 & 0 & 0 & 0 & 0 \end{bmatrix}$$
Transition probability matrix at 1 year:
$$P(1) = \begin{bmatrix} .734 & .214 & .025 & .027 \\ .021 & .741 & .179 & .059 \\ 0 & 0 & 1 & 0 \\ 0 & 0 & 0 & 1 \end{bmatrix}$$

Mean time spent by the susceptible individuals in state 1 is approximately 3 years and 2 months, and in state 2 the mean sojourn time is approximately 3 years and 3.5 months .According to American Association for the study of Liver Disease[8],the most common cause of death in patients with NAFLD is cardiovascular disease (CVD) independent of other metabolic comorbidities , whether the liver-related mortality is the second or third cause of death among patients with NAFLD. Cancer-related mortality is among the top three causes of death in subjects with NAFLD. As shown from the calculations; mean time to absorption can be classified into : mean time from state 1( susceptible individuals with risk factors) to state 3 ( liver-related mortality)is approximately 5 years, while the mean time from state 1 to state 4 ( for example CVD as an example for causes of death other than liver-related mortality causes) is approximately 2 years .The mean time from state 2( NAFLD) to state 3 ( liver-related mortality ) is approximately 3 years while it decreases to approximately 1 year from state 2 ( NAFLD) to state 4 ( other causes than liver-related mortality).

If a cohort of 3000 susceptible individuals have initial distribution of $[.7 \quad .3 \quad 0 \quad 0]$, and initial number of individuals in each state $[2100 \quad 900 \quad 0 \quad 0]$, then at 1 year the state probability distribution is $[.52 \quad .372 \quad .071 \quad .037]$ and the expected counts of patients at each state are $[1559 \quad 1117 \quad 214 \quad 110]$

But at 60 years the state probability distribution is $[0 \quad 0 \quad .715 \quad .285]$ and the expected counts of patients at each state are $[0 \quad 0 \quad 2145 \quad 855]$ while the asymptotic covariance matrix for the state probability distribution is
$$\begin{bmatrix} .088589 & .16401 & 0 & 0 \\ .16401 & .30368 & 0 & 0 \\ 0 & 0 & 0 & 0 \\ 0 & 0 & 0 & 0 \end{bmatrix}$$

To calculate goodness of fit for multistate model used in the small model, it is like the procedure used in contingency table, and it is calculated in each interval then sum up:

Step 1 :
$H_0 = $ future state does not depend on the current state.
$H_1 = $ future state does depend on the current state
Step 2: calculate the
$$P_{ij}(\Delta t = 1) = \begin{bmatrix} .7338 & .2139 & .0247 & .0277 \\ .0206 & .7411 & .1793 & .059 \\ 0 & 0 & 1 & 0 \\ 0 & 0 & 0 & 1 \end{bmatrix}$$
by exponentiation of the estimated Q matrix
step 3 :calculate the expected counts in this interval by multiplying each row in the probability matrix with the corresponding total marginal counts in the observed transition counts matrix in the same interval to get the expected counts.

|  | State 1 | State 2 | State3 | State4 | total |
|---|---|---|---|---|---|
| State1 | 403.59 | 117.645 | 13.585 | 15.235 | 550.055 |
| State2 | 5.15 | 185.275 | 44.825 | 14.75 | 250 |
| State3 | 0 | 0 | 0 | 0 | 0 |
| State4 | 0 | 0 | 0 | 0 | 0 |

Step 4: apply
$$\sum_{i=1}^{4} \sum_{j=1}^{4} \frac{(O_{ij} - E_{ij})^2}{E_{ij}} = 104.247 \sim \chi^2_{(4-1)(4-1)(.05)}$$
The same steps are used for the observed transition counts in the $\Delta t = 2$ and $\Delta t = 3$ with the following results:
$$P_{ij}(\Delta t = 2) = \begin{bmatrix} .5428 & .3154 & .0811 & .0607 \\ .0304 & .5537 & .3126 & .1033 \\ 0 & 0 & 1 & 0 \\ 0 & 0 & 0 & 1 \end{bmatrix}$$

The expected counts:

|  | State 1 | State 2 | State3 | State4 | total |
|---|---|---|---|---|---|
| State1 | 60.2508 | 35.0094 | 9.0021 | 6.7377 | 111 |
| State2 | 1.1856 | 21.5943 | 12.1914 | 4.0287 | 39 |
| State3 | 0 | 0 | 0 | 0 | 0 |
| State4 | 0 | 0 | 0 | 0 | 0 |

$$\sum_{i=1}^{4}\sum_{j=1}^{4}\frac{(O_{ij}-E_{ij})^2}{E_{ij}} = 8.022 \sim \chi^2_{(4-1)(4-1)(.05)}$$

The same steps are used for the observed transition counts in $\Delta t = 3$ with the following results:

$$P_{ij}(\Delta t = 3) = \begin{bmatrix} .4048 & .3499 & .151 & .0943 \\ .0337 & .4168 & .4126 & .1368 \\ 0 & 0 & 1 & 0 \\ 0 & 0 & 0 & 1 \end{bmatrix}$$

The expected counts:

|  | State 1 | State 2 | State3 | State4 | total |
|---|---|---|---|---|---|
| State1 | 15.7872 | 13.6461 | 5.889 | 3.6777 | 39 |
| State2 | .3707 | 4.5848 | 4.5386 | 1.5048 | 11 |
| State3 | 0 | 0 | 0 | 0 | 0 |
| State4 | 0 | 0 | 0 | 0 | 0 |

$$\sum_{i=1}^{4}\sum_{j=1}^{4}\frac{(O_{ij}-E_{ij})^2}{E_{ij}} = 6.588 \sim \chi^2_{(4-1)(4-1)(.05)}$$

Step 5: sum up the above results to get:

$$\sum_{i=1}^{4}\sum_{j=1}^{4}\sum_{l=1}^{t=3}\frac{(O_{ijl}-E_{ijl})^2}{E_{ijl}} = 118.857 \sim \chi^2_{(df=27)(.05)}$$

So from the above results the null hypothesis is rejected while the alternative hypothesis is accepted and the model fits the data that is to mean the future state depends on the current state with the estimated transition rate and probability matrices as obtained.

## VII. Summary and conclusion

Nonalcoholic fatty liver disease is one of the most common causes of liver disease worldwide. Understanding natural history of NAFLD is mandatory to calculate and predict future clinical outcome and economic burden used to improve the diagnostic utilities and tools of the disease as well as therapeutic procedures. This is accomplished by developing statistical models that offer these calculations to health care providers and health policy makers to design plans that confront these challenges in management of this disease process aiming to ameliorate its progression and complications. An example of the non-invasive diagnostic tools is the circulating level of cytokeratin-18 fragments, although promising it is not available in a clinical care setting and there is not an established cut-off value for identifying steato-hepatitis (NASH)[20]. A genetic polymorphism of patatin-like phospholipase domain-containing protein 3 gene variants (PNPLA-3) are associated with NASH and advanced fibrosis, however testing for these variants in routine clinical care is not supported. More studies may be of longitudinal orientation, like multistate Markov models may be required to attain more research evidence base validation for their use in routine clinical setting.

Multistate Markov chains are one of most frequently used and great potentiality offering models for such analysis. These chains can be used compactly as in this paper describing the disease in its simplest form as well as they can be used by expanding the disease states in more detailed form that describes the disease process in more informative stages each represented by a specific well defined criteria for each state. Other models such as hidden Markov chains and semi-Markov chains can provide more statistical information to the health care policy makers for better management.

**Abbreviations :**
CTMC: continuous time Markov chains, CVS: cardiovascular disease, EASD: European Association for the Study of diabetes, EASL: European Association for the Study of liver, EASO: European Association for the Study of obesity, HS: hepatic steatosis, NAFLD: non-alcoholic fatty liver disease, NASH: non-alcoholic steatohepatitis, PNPLA-3:patatin-like phospholipase domain-containing protein 3 gene variants, TE: transient elastography, T2DM: type 2 diabetes mellitus.

**Declarations:**
**Ethics approval and consent to participate**
Not applicable.
**Consent for publication**
Not applicable
**Availability of data and material**
Not applicable. Data sharing not applicable to this article as no datasets were generated or analyzed during the current study.
**Competing interests**
The author declares that I have no competing interests.
**Funding**
No funding resource. No funding roles in the design of the study and collection, analysis, and interpretation of data and in writing the manuscript are declared
**Authors' contribution**
I am the author who has carried the mathematical analysis as well as applying these mathematical statistical concepts on the hypothetical example.
**Acknowledgement**
Not applicable.

# Appendix
## 1. Transition Rates And Probabilities

$$\frac{d}{dt}P_{ij}(t)$$

$$= \begin{bmatrix} P_{11} & P_{12} & P_{13} & P_{14} \\ P_{21} & P_{22} & P_{23} & P_{24} \\ 0 & 0 & P_{33} & 0 \\ 0 & 0 & 0 & P_{44} \end{bmatrix} \begin{bmatrix} -(\lambda_{12} + \lambda_{14}) & \lambda_{12} & 0 & \lambda_{14} \\ \mu_{21} & -(\mu_{21} + \lambda_{23} + \lambda_{24}) & \lambda_{23} & \lambda_{24} \\ 0 & 0 & 0 & 0 \\ 0 & 0 & 0 & 0 \end{bmatrix}$$

The Kolmogrove differential equations:

$$\frac{dP_{11}}{dt} = -P_{11}(\lambda_{12} + \lambda_{14}) + P_{12}\mu_{21}, \frac{dP_{12}}{dt}$$
$$= P_{11}\lambda_{12} - P_{12}(\mu_{21} + \lambda_{23} + \lambda_{24})$$

$$\frac{dP_{13}}{dt} = P_{12}\lambda_{23}, \frac{dP_{14}}{dt} = P_{11}\lambda_{14} + P_{12}\lambda_{24}$$

$$\frac{dP_{21}}{dt} = -P_{21}(\lambda_{12} + \lambda_{14}) + P_{22}\mu_{21}, \frac{dP_{22}}{dt}$$
$$= P_{21}\lambda_{12} - P_{22}(\mu_{21} + \lambda_{23} + \lambda_{24})$$

$$\frac{dP_{23}}{dt} = P_{22}\lambda_{23}, \frac{dP_{24}}{dt} = P_{21}\lambda_{14} + P_{22}\lambda_{24}$$

$P_{33} = 1$, $P_{44} = 1$

This is a system of differential equations and the followings are the solutions for its components:

To solve the set of probabilities in the first row:
The first 2 equations are:

$$\frac{dP_{11}(t)}{dt} = -P_{11}(\lambda_{12} + \lambda_{14}) + P_{12}\mu_{21}$$

$$\frac{dP_{12}(t)}{dt} = P_{11}\lambda_{12} - P_{12}(\mu_{21} + \lambda_{23} + \lambda_{24})$$

let $\lambda_{12} + \lambda_{14} = \gamma_1$
let $\mu_{21} + \lambda_{23} + \lambda_{24} = \gamma_2$

To get $P_{11}$

$DP_{11} + \gamma_1 P_{11} - \mu_{21} P_{12} = 0$     (1)
$DP_{12} + \gamma_2 P_{12} - \lambda_{12} P_{11} = 0$     (2)
$(D + \gamma_1) P_{11} - \mu_{21} P_{12} = 0$     (3)
$-\lambda_{12} P_{11} + (D + \gamma_2) P_{12} = 0$     (4)

Multiply equation (3) by $(D + \gamma_2)$ and multiply equation (4) by $\mu_{21}$

$(D + \gamma_1)(D + \gamma_2) P_{11} - (D + \gamma_2)\mu_{21} P_{12} = 0$     (5)
$-\lambda_{12} \mu_{21} P_{11} + (D + \gamma_2)\mu_{21} P_{12} = 0$     (6)

Add the above equations:
$[(D + \gamma_1)(D + \gamma_2) - \lambda_{12}\mu_{21}] P_{11} = 0$
$[D^2 + (\gamma_1 + \gamma_2)D + \gamma_1\gamma_2 - \lambda_{12}\mu_{21}] P_{11} = 0$

$$w_1 = \frac{-(\gamma_1 + \gamma_2) - \sqrt{(\gamma_1 + \gamma_2)^2 - 4\gamma_1\gamma_2 + 4\lambda_{12}\mu_{21}}}{2}$$

$$w_2 = \frac{-(\gamma_1 + \gamma_2) + \sqrt{(\gamma_1 + \gamma_2)^2 - 4\gamma_1\gamma_2 + 4\lambda_{12}\mu_{21}}}{2}$$

$(\gamma_1 + \gamma_2)^2 - 4\gamma_1\gamma_2 + 4\lambda_{12}\mu_{21} > 0$
$P_{11} = c_1 e^{w_1 t} + c_2 e^{w_2 t}$

To get $P_{12}$
$(D + \gamma_1) P_{11} - \mu_{21} P_{12} = 0$     (3)
$-\lambda_{12} P_{11} + (D + \gamma_2) P_{12} = 0$     (4)

Multiply equation (3) by $\lambda_{12}$ and multiply equation (4) by $(D + \gamma_1)$

$(D + \gamma_1) \lambda_{12} P_{11} - \lambda_{12}\mu_{21} P_{12} = 0$     (5)
$-(D + \gamma_1) \lambda_{12} P_{11} + (D + \gamma_1)(D + \gamma_2) P_{12} = 0$     (6)

Add the above equations:
$[(D + \gamma_1)(D + \gamma_2) - \lambda_{12}\mu_{21}] P_{12} = 0$
$[D^2 + (\gamma_1 + \gamma_2)D + \gamma_1\gamma_2 - \lambda_{12}\mu_{21}] P_{12} = 0$

$$w_1 = \frac{-(\gamma_1 + \gamma_2) - \sqrt{(\gamma_1 + \gamma_2)^2 - 4\gamma_1\gamma_2 + 4\lambda_{12}\mu_{21}}}{2}$$

$$w_2 = \frac{-(\gamma_1 + \gamma_2) + \sqrt{(\gamma_1 + \gamma_2)^2 - 4\gamma_1\gamma_2 + 4\lambda_{12}\mu_{21}}}{2}$$

$(\gamma_1 + \gamma_2)^2 - 4\gamma_1\gamma_2 + 4\lambda_{12}\mu_{21} > 0$
$P_{12} = c_3 e^{w_1 t} + c_4 e^{w_2 t}$

Substitute in:
$DP_{11} + \gamma_1 P_{11} - \mu_{21} P_{12} = 0$
$c_1 w_1 e^{w_1 t} + c_2 w_2 e^{w_2 t} + c_1 \gamma_1 e^{w_1 t} + c_2 \gamma_1 e^{w_2 t} - \mu_{21} c_3 e^{w_1 t} - \mu_{21} c_4 e^{w_2 t} = 0$
$(c_1 w_1 + c_1 \gamma_1 - \mu_{21} c_3) e^{w_1 t} + (c_2 w_2 + c_2 \gamma_1 - \mu_{21} c_4) e^{w_2 t} = 0$

| $c_1 w_1 + c_1 \gamma_1 - \mu_{21} c_3 = 0$ | $c_2 w_2 + c_2 \gamma_1 - \mu_{21} c_4 = 0$ |
|---|---|
| $\mu_{21} c_3 = c_1 w_1 + c_1 \gamma_1$ | $\mu_{21} c_4 = c_2 w_2 + c_2 \gamma_1$ |
| $c_3 = \dfrac{c_1}{\mu_{21}}(w_1 + \gamma_1)$ | $c_4 = \dfrac{c_2}{\mu_{21}}(w_2 + \gamma_1)$ |

$P_{11} = c_1 e^{w_1 t} + c_2 e^{w_2 t}$
$P_{12} = c_3 e^{w_1 t} + c_4 e^{w_2 t} = \dfrac{c_1}{\mu_{21}}(w_1 + \gamma_1) e^{w_1 t} + \dfrac{c_2}{\mu_{21}}(w_2 + \gamma_1) e^{w_2 t}$

Using initial values at:

| $P_{11}(0) = 1$ | $P_{12}(0) = 0$ |
|---|---|
| $c_1 + c_2 = 1$ | $\dfrac{c_1}{\mu_{21}}(w_1 + \gamma_1) + \dfrac{c_2}{\mu_{21}}(w_2 + \gamma_1) = 0$ |
| $c_1 = 1 - c_2$ | |
| | $c_1(w_1 + \gamma_1) + c_2(w_2 + \gamma_1) = 0$ |
| | $c_1 w_1 + c_1 \gamma_1 + c_2 w_2 + c_2 \gamma_1 = 0$ |
| | $c_1 w_1 + c_2 w_2 + (c_2 + c_1)\gamma_1 = 0$ |
| | $c_1 w_1 + c_2 w_2 + \gamma_1 = 0$ |
| | $c_1 w_1 + c_2 w_2 = -\gamma_1$ |

$(1 - c_2) w_1 + c_2 w_2 = -\gamma_1$
$w_1 - c_2 w_1 + c_2 w_2 = -\gamma_1$
$-c_2 w_1 + c_2 w_2 = -\gamma_1 - w_1 = -(\gamma_1 + w_1)$
$c_2 (w_2 - w_1) = -\gamma_1 - w_1 = -(\gamma_1 + w_1)$

$$c_2 = \left(\frac{w_1 + \gamma_1}{w_1 - w_2}\right)$$

$$c_1 = (1 - c_2) = 1 - \frac{(\gamma_1 + w_1)}{(w_1 - w_2)} = \left(\frac{w_2 + \gamma_1}{w_2 - w_1}\right)$$

$$P_{11} = \left(\frac{w_2 + \gamma_1}{w_2 - w_1}\right) e^{w_1 t} + \left(\frac{w_1 + \gamma_1}{w_1 - w_2}\right) e^{w_2 t}$$

$$P_{12} = \frac{c_1}{\mu_{21}}(w_1 + \gamma_1) e^{w_1 t} + \frac{c_2}{\mu_{21}}(w_2 + \gamma_1) e^{w_2 t}$$

$$P_{12} = \left(\frac{w_2 + \gamma_1}{w_2 - w_1}\right)\left(\frac{w_1 + \gamma_1}{\mu_{21}}\right) e^{w_1 t} + \left(\frac{w_1 + \gamma_1}{w_1 - w_2}\right)\left(\frac{w_2 + \gamma_1}{\mu_{21}}\right) e^{w_2 t}$$

let: $\left(\dfrac{w_2 + \gamma_1}{w_2 - w_1}\right) = A_1$, $\left(\dfrac{w_1 + \gamma_1}{w_1 - w_2}\right) = A_2$

$\left(\dfrac{w_2 + \gamma_1}{w_2 - w_1}\right)\left(\dfrac{w_1 + \gamma_1}{\mu_{21}}\right) = A_3$, $\left(\dfrac{w_1 + \gamma_1}{w_1 - w_2}\right)\left(\dfrac{w_2 + \gamma_1}{\mu_{21}}\right) = A_4$

$\therefore P_{11} = A_1 e^{w_1 t} + A_2 e^{w_2 t}$
$\therefore P_{12} = A_3 e^{w_1 t} + A_4 e^{w_2 t}$

To get $P_{13}$

$$\frac{dP_{13}}{dt} = P_{12}\lambda_{23}$$

$$\frac{dP_{13}}{dt} = \lambda_{23}(A_3 e^{w_1 t} + A_4 e^{w_2 t}) = \lambda_{23} A_3 e^{w_1 t} + \lambda_{23} A_4 e^{w_2 t}$$

$$P_{13} = \left[\lambda_{23} A_3 \frac{e^{w_1 t}}{w_1} - \frac{\lambda_{23} A_3}{w_1}\right] + \left[\lambda_{23} A_4 \frac{e^{w_2 t}}{w_2} - \frac{\lambda_{23} A_4}{w_2}\right]$$

$$P_{13} = \frac{\lambda_{23} A_3}{w_1}(e^{w_1 t} - 1) + \frac{\lambda_{23} A_4}{w_2}(e^{w_2 t} - 1)$$

let $\dfrac{\lambda_{23} A_3}{w_1} = A_5$, $\dfrac{\lambda_{23} A_4}{w_2} = A_6$

$\therefore P_{13} = A_5 (e^{w_1 t} - 1) + A_6 (e^{w_2 t} - 1)$

To get $P_{14}$

$$\frac{dP_{14}}{dt} = \lambda_{14}(A_1 e^{w_1 t} + A_2 e^{w_2 t}) + \lambda_{24}(A_3 e^{w_1 t} + A_4 e^{w_2 t})$$

$$\frac{dP_{14}}{dt} = \lambda_{14} A_1 e^{w_1 t} + \lambda_{14} A_2 e^{w_2 t} + \lambda_{24} A_3 e^{w_1 t} + \lambda_{24} A_4 e^{w_2 t}$$

$$\frac{dP_{14}}{dt} = (\lambda_{14} A_1 + \lambda_{24} A_3) e^{w_1 t} + (\lambda_{14} A_2 + \lambda_{24} A_4) e^{w_2 t}$$

let: $(\lambda_{14} A_1 + \lambda_{24} A_3) = G_1$, $(\lambda_{14} A_2 + \lambda_{24} A_4) = G_2$

$$\frac{dP_{14}}{dt} = G_1 e^{w_1 t} + G_2 e^{w_2 t}$$

$$P_{14} = \left[G_1 \frac{e^{w_1 t}}{w_1} - \frac{G_1}{w_1}\right] + \left[G_2 \frac{e^{w_2 t}}{w_2} - \frac{G_2}{w_2}\right]$$

$$= \frac{G_1}{w_1}(e^{w_1 t} - 1) + \frac{G_2}{w_2}(e^{w_2 t} - 1)$$

let: $\dfrac{G_1}{w_1} = A_7$, $\dfrac{G_2}{w_2} = A_8$

$\therefore P_{14} = A_7(e^{w_1 t} - 1) + A_8(e^{w_2 t} - 1)$

To solve the set of probabilities in the second row:

$$\frac{dP_{21}}{dt} = -P_{21}(\lambda_{12} + \lambda_{14}) + P_{22}\mu_{21}$$

$$\frac{dP_{22}}{dt} = P_{21}\lambda_{12} - P_{22}(\mu_{21} + \lambda_{23} + \lambda_{24})$$

previously $\lambda_{12} + \lambda_{14} = \gamma_1$ and $\mu_{21} + \lambda_{23} + \lambda_{24} = \gamma_2$

*To get $P_{21}$*
$DP_{21} + \gamma_1 P_{21} - \mu_{21} P_{22} = 0$     (1)
$DP_{22} + \gamma_2 P_{22} - \lambda_{12} P_{21} = 0$     (2)
$(D + \gamma_1) P_{21} - \mu_{21} P_{22} = 0$     (3)
$-\lambda_{12} P_{21} + (D + \gamma_2) P_{22} = 0$     (4)

*Multiply equation (3) by $(D + \gamma_2)$ and multiply equation (4) by $\mu_{21}$*
$(D + \gamma_1)(D + \gamma_2) P_{21} - (D + \gamma_2) \mu_{21} P_{22} = 0$    (5)
$-\lambda_{12} \mu_{21} P_{21} + (D + \gamma_2) \mu_{21} P_{22} = 0$     (6)

Add the above equations :
$[(D + \gamma_1)(D + \gamma_2) - \lambda_{12}\mu_{21}] P_{21} = 0$
$[D^2 + (\gamma_1 + \gamma_2)D + \gamma_1 \gamma_2 - \lambda_{12}\mu_{21}] P_{21} = 0$

$w_1 = \dfrac{-(\gamma_1 + \gamma_2) - \sqrt{(\gamma_1 + \gamma_2)^2 - 4\gamma_1\gamma_2 + 4\lambda_{12}\mu_{21}}}{2}$

$w_2 = \dfrac{-(\gamma_1 + \gamma_2) + \sqrt{(\gamma_1 + \gamma_2)^2 - 4\gamma_1\gamma_2 + 4\lambda_{12}\mu_{21}}}{2}$

$(\gamma_1 + \gamma_2)^2 - 4\gamma_1\gamma_2 + 4\lambda_{12}\mu_{21} > 0$
$P_{21} = c_5 e^{w_1 t} + c_6 e^{w_2 t}$

*To get $P_{22}$*
$(D + \gamma_1) P_{21} - \mu_{21} P_{22} = 0$     (3)
$-\lambda_{12} P_{21} + (D + \gamma_2) P_{22} = 0$     (4)

*Multiply equation (3) by $\lambda_{12}$ and multiply equation (4) by $(D + \gamma_1)$*
$(D + \gamma_1) \lambda_{12} P_{21} - \lambda_{12} \mu_{21} P_{22} = 0$    (5)
$-(D + \gamma_1) \lambda_{12} P_{21} + (D + \gamma_1)(D + \gamma_2) P_{22} = 0$    (6)

Add the above equations:
$[(D + \gamma_1)(D + \gamma_2) - \lambda_{12}\mu_{21}] P_{22} = 0$
$[D^2 + (\gamma_1 + \gamma_2)D + \gamma_1\gamma_2 - \lambda_{12}\mu_{21}] P_{22} = 0$

$w_1 = \dfrac{-(\gamma_1 + \gamma_2) - \sqrt{(\gamma_1 + \gamma_2)^2 - 4\gamma_1\gamma_2 + 4\lambda_{12}\mu_{21}}}{2}$

$w_2 = \dfrac{-(\gamma_1 + \gamma_2) + \sqrt{(\gamma_1 + \gamma_2)^2 - 4\gamma_1\gamma_2 + 4\lambda_{12}\mu_{21}}}{2}$

$(\gamma_1 + \gamma_2)^2 - 4\gamma_1\gamma_2 + 4\lambda_{12}\mu_{21} > 0$
$P_{22} = c_7 e^{w_1 t} + c_8 e^{w_2 t}$

Substitute in:
$DP_{21} + \gamma_1 P_{21} - \mu_{21} P_{22} = 0$
$c_5 w_1 e^{w_1 t} + c_6 w_2 e^{w_2 t} + c_5 \gamma_1 e^{w_1 t} + c_6 \gamma_1 e^{w_2 t} - \mu_{21} c_7 e^{w_1 t} - \mu_{21} c_8 e^{w_2 t} = 0$
$(c_5 w_1 + c_5 \gamma_1 - \mu_{21} c_7) e^{w_1 t} + (c_6 w_2 + c_6 \gamma_1 - \mu_{21} c_8) e^{w_2 t} = 0$

| $c_5 w_1 + c_5 \gamma_1 - \mu_{21} c_7 = 0$ | $c_6 w_2 + c_6 \gamma_1 - \mu_{21} c_8 = 0$ |
|---|---|
| $\mu_{21} c_7 = c_5 w_1 + c_5 \gamma_1$ | $\mu_{21} c_8 = c_6 w_2 + c_6 \gamma_1$ |
| $c_7 = \dfrac{c_5}{\mu_{21}}(w_1 + \gamma_1)$ | $c_8 = \dfrac{c_6}{\mu_{21}}(w_2 + \gamma_1)$ |

$P_{21} = c_5 e^{w_1 t} + c_6 e^{w_2 t}$
$P_{22} = c_7 e^{w_1 t} + c_8 e^{w_2 t} = \dfrac{c_5}{\mu_{21}}(w_1 + \gamma_1) e^{w_1 t} + \dfrac{c_6}{\mu_{21}}(w_2 + \gamma_1) e^{w_2 t}$

Using initial values at :

| $P_{21}(0) = 0$ | $P_{22}(0) = 1$ |
|---|---|
| $c_5 + c_6 = 0$ | $\dfrac{c_5}{\mu_{21}}(w_1 + \gamma_1) + \dfrac{c_6}{\mu_{21}}(w_2 + \gamma_1) = 1$ |
| $c_5 = -c_6$ | $c_5(w_1 + \gamma_1) + c_6(w_2 + \gamma_1) = \mu_{21}$ |
| | $c_5 w_1 + c_5 \gamma_1 + c_6 w_2 + c_6 \gamma_1 = \mu_{21}$ |
| | $c_5 w_1 + c_6 w_2 + (c_5 + c_6)\gamma_1 = \mu_{21}$ |
| | $c_5 w_1 + c_6 w_2 = \mu_{21}$ |
| | $-c_6 w_1 + c_6 w_2 = \mu_{21}$ |
| | $c_6 = \dfrac{\mu_{21}}{(w_2 - w_1)}$ , $c_5 = \dfrac{\mu_{21}}{(w_1 - w_2)}$ |

$P_{21} = c_5 e^{w_1 t} + c_6 e^{w_2 t}$
$P_{21} = \left(\dfrac{\mu_{21}}{w_1 - w_2}\right) e^{w_1 t} + \left(\dfrac{\mu_{21}}{w_2 - w_1}\right) e^{w_2 t} = \left(\dfrac{\mu_{21}}{w_1 - w_2}\right)(e^{w_1 t} - e^{w_2 t})$

$P_{22} = c_7 e^{w_1 t} + c_8 e^{w_2 t} = \dfrac{c_5}{\mu_{21}}(w_1 + \gamma_1) e^{w_1 t} + \dfrac{c_6}{\mu_{21}}(w_2 + \gamma_1) e^{w_2 t}$

$P_{22} = c_7 e^{w_1 t} + c_8 e^{w_2 t} = \left(\dfrac{\mu_{21}}{w_1 - w_2}\right)\left(\dfrac{w_1 + \gamma_1}{\mu_{21}}\right) e^{w_1 t} + \left(\dfrac{\mu_{21}}{w_2 - w_1}\right)\left(\dfrac{w_2 + \gamma_1}{\mu_{21}}\right) e^{w_2 t}$

$P_{22} = \left(\dfrac{w_1 + \gamma_1}{w_1 - w_2}\right) e^{w_1 t} + \left(\dfrac{w_2 + \gamma_1}{w_2 - w_1}\right) e^{w_2 t}$

let $\left(\dfrac{\mu_{21}}{w_1 - w_2}\right) = A_9$ , $\left(\dfrac{w_1 + \gamma_1}{w_1 - w_2}\right) = A_{10}$ , $\left(\dfrac{w_2 + \gamma_1}{w_2 - w_1}\right) = A_{11}$

$\therefore P_{21} = A_9 (e^{w_1 t} - e^{w_2 t})$
$\therefore P_{22} = A_{10} e^{w_1 t} + A_{11} e^{w_2 t}$

*To get $P_{23}$*
$\dfrac{dP_{23}}{dt} = P_{22} \lambda_{23}$

$\dfrac{dP_{23}}{dt} = \lambda_{23}(A_{10} e^{w_1 t} + A_{11} e^{w_2 t}) = \lambda_{23} A_{10} e^{w_1 t} + \lambda_{23} A_{11} e^{w_2 t}$

$P_{23} = \left[\lambda_{23} A_{10} \dfrac{e^{w_1 t}}{w_1} - \dfrac{\lambda_{23} A_{10}}{w_1}\right] + \left[\lambda_{23} A_{11} \dfrac{e^{w_2 t}}{w_2} - \dfrac{\lambda_{23} A_{11}}{w_2}\right]$

$P_{23} = \dfrac{\lambda_{23} A_{10}}{w_1}(e^{w_1 t} - 1) + \dfrac{\lambda_{23} A_{11}}{w_2}(e^{w_2 t} - 1)$

let $\dfrac{\lambda_{23} A_{10}}{w_1} = A_{12}$ , $\dfrac{\lambda_{23} A_{11}}{w_2} = A_{13}$

$\therefore P_{23} = A_{12}(e^{w_1 t} - 1) + A_{13}(e^{w_2 t} - 1)$

*To get $P_{24}$*
$\dfrac{dP_{24}}{dt} = P_{21}\lambda_{14} + P_{22}\lambda_{24}$

$\dfrac{dP_{24}}{dt} = \lambda_{14} A_9(e^{w_1 t} - e^{w_2 t}) + \lambda_{24}(A_{10} e^{w_1 t} + A_{11} e^{w_2 t})$

$\dfrac{dP_{24}}{dt} = \lambda_{14} A_9 e^{w_1 t} - \lambda_{14} A_9 e^{w_2 t} + \lambda_{24} A_{10} e^{w_1 t} + \lambda_{24} A_{11} e^{w_2 t}$

$\dfrac{dP_{24}}{dt} = (\lambda_{14} A_9 + \lambda_{24} A_{10}) e^{w_1 t} + (\lambda_{24} A_{11} - \lambda_{14} A_9) e^{w_2 t}$

let : $(\lambda_{14} A_9 + \lambda_{24} A_{10}) = G_3$ , $(\lambda_{24} A_{11} - \lambda_{14} A_9) = G_4$

$\dfrac{dP_{24}}{dt} = G_3 e^{w_1 t} + G_4 e^{w_2 t}$

$P_{24} = \left[G_3 \dfrac{e^{w_1 t}}{w_1} - \dfrac{G_3}{w_1}\right] + \left[G_4 \dfrac{e^{w_2 t}}{w_2} - \dfrac{G_4}{w_2}\right]$

$= \dfrac{G_3}{w_1}(e^{w_1 t} - 1) + \dfrac{G_4}{w_2}(e^{w_2 t} - 1)$

let : $\dfrac{G_3}{w_1} = A_{14}$ , $\dfrac{G_4}{w_2} = A_{15}$

$\therefore P_{24} = A_{14}(e^{w_1 t} - 1) + A_{15}(e^{w_2 t} - 1)$

$P_{33} = 1$
$P_{44} = 1$

**A. MLE to Estimate Transition Rate Matrix**

$Q = \begin{bmatrix} -(\lambda_{12} + \lambda_{14}) & \lambda_{12} & 0 & \lambda_{14} \\ \mu_{21} & -(\mu_{21} + \lambda_{23} + \lambda_{24}) & \lambda_{23} & \lambda_{24} \\ 0 & 0 & 0 & 0 \\ 0 & 0 & 0 & 0 \end{bmatrix}$

let $(\lambda_{12} + \lambda_{14}) = \gamma_1$ , $(\mu_{21} + \lambda_{23} + \lambda_{24}) = \gamma_2$

$\dfrac{\partial}{\partial \theta} P_{ij}(t)$
$= t e^{\Lambda t} d\Lambda$ , where $\Lambda$ are the eiqenvalues of the Q matrix .
The eigenvalues are the solution of the $|Q - \rho I| = 0$
and the $|Q - \rho I|$ gives the characteristic polynomial of this matrix :
$[-\lambda_{12}\mu_{21} + (\gamma_1 + \rho)(\gamma_2 + \rho)](-\rho)(-\rho) = 0$
$[-\lambda_{12}\mu_{21} + (\gamma_1 + \rho)(\gamma_2 + \rho)](-\rho)(-\rho)$
$\quad\quad\quad\quad = \rho^2 [\rho^2 + (\gamma_1 + \gamma_2)\rho + \gamma_1 \gamma_2 - \lambda_{12}\mu_{21}] = 0$

$\rho = \{0, 0, \rho_3, \rho_4\}$

$\rho_3 = \dfrac{-(\gamma_1 + \gamma_2) - \sqrt{[(\gamma_1 + \gamma_2)]^2 - 4\gamma_1\gamma_2 + 4\mu_{21}\lambda_{12}}}{2}$

$\quad = \dfrac{-(\gamma_1 + \gamma_2) - \sqrt{.}}{2}$

$\rho_4 = \dfrac{-(\gamma_1 + \gamma_2) + \sqrt{[(\gamma_1 + \gamma_2)]^2 - 4\gamma_1\gamma_2 + 4\mu_{21}\lambda_{12}}}{2}$

$\quad = \dfrac{-(\gamma_1 + \gamma_2) + \sqrt{.}}{2}$

$[(\gamma_1 + \gamma_2)]^2 - 4\gamma_1\gamma_2 + 4\mu_{21}\lambda_{12} = \sqrt{.} =$
$(\lambda_{12}^2 + \lambda_{14}^2 + \mu_{21}^2 + \lambda_{23}^2 + \lambda_{24}^2 + 2\lambda_{12}\lambda_{14} + 2\lambda_{12}\mu_{21} + 2\lambda_{23}\mu_{21} + 2\lambda_{24}\mu_{21}$
$\quad\quad + 2\lambda_{23}\lambda_{24} - 2\lambda_{12}\lambda_{23} - 2\lambda_{12}\lambda_{24} - 2\lambda_{14}\mu_{21} - 2\lambda_{14}\lambda_{23}$
$\quad\quad - 2\lambda_{14}\lambda_{24})^{.5}$

$\rho_3 = \dfrac{1}{2}\{-\lambda_{12} - \lambda_{14} - \mu_{21} - \lambda_{23} - \lambda_{24}$
$\quad\quad - (\lambda_{12}^2 + \lambda_{14}^2 + \mu_{21}^2 + \lambda_{23}^2 + \lambda_{24}^2 + 2\lambda_{12}\lambda_{14} + 2\lambda_{12}\mu_{21}$
$\quad\quad + 2\lambda_{23}\mu_{21} + 2\lambda_{24}\mu_{21} + 2\lambda_{23}\lambda_{24} - 2\lambda_{12}\lambda_{23} - 2\lambda_{12}\lambda_{24}$
$\quad\quad - 2\lambda_{14}\mu_{21} - 2\lambda_{14}\lambda_{23} - 2\lambda_{14}\lambda_{24})^{.5}\}$

$$\rho_4 = \frac{1}{2}\{-\lambda_{12} - \lambda_{14} - \mu_{21} - \lambda_{23} - \lambda_{24}$$
$$+ (\lambda_{12}^2 + \lambda_{14}^2 + \mu_{21}^2 + \lambda_{23}^2 + \lambda_{24}^2 + 2\lambda_{12}\lambda_{14} + 2\lambda_{12}\mu_{21}$$
$$+ 2\lambda_{23}\mu_{21} + 2\lambda_{24}\mu_{21} + 2\lambda_{23}\lambda_{24} - 2\lambda_{12}\lambda_{23} - 2\lambda_{12}\lambda_{24}$$
$$- 2\lambda_{14}\mu_{21} - 2\lambda_{14}\lambda_{23} - 2\lambda_{14}\lambda_{24})^{.5}\}$$

$$\theta_1 = \lambda_{12}, \quad \theta_2 = \lambda_{14}, \quad \theta_3 = \mu_{21}, \quad \theta_4 = \lambda_{23}, \quad \theta_5 = \lambda_{24}$$

$$\frac{\partial}{\partial \lambda_{12}}\rho_3 = -\frac{1}{2} - \frac{1}{2}\left(\frac{1}{2}\right)(.)^{-.5}(2\lambda_{12} + 2\lambda_{14} + 2\mu_{21} - 2\lambda_{23} - 2\lambda_{24})$$
$$= -\frac{1}{2} - \frac{1}{2}(.)^{-.5}(\lambda_{12} + \lambda_{14} + \mu_{21} - \lambda_{23} - \lambda_{24})$$

$$\frac{\partial}{\partial \lambda_{14}}\rho_3 = -\frac{1}{2} - \frac{1}{2}\left(\frac{1}{2}\right)(.)^{-.5}(2\lambda_{14} + 2\lambda_{12} - 2\mu_{21} - 2\lambda_{23} - 2\lambda_{24})$$
$$= -\frac{1}{2} - \frac{1}{2}(.)^{-.5}(\lambda_{12} + \lambda_{14} - \mu_{21} - \lambda_{23} - \lambda_{24})$$

$$\frac{\partial}{\partial \mu_{21}}\rho_3 = -\frac{1}{2} - \frac{1}{2}\left(\frac{1}{2}\right)(.)^{-.5}(2\mu_{21} + 2\lambda_{12} + 2\lambda_{23} + 2\lambda_{24} - 2\lambda_{14})$$
$$= -\frac{1}{2} - \frac{1}{2}(.)^{-.5}(\mu_{21} + \lambda_{12} + \lambda_{23} + \lambda_{24} - \lambda_{14})$$

$$\frac{\partial}{\partial \lambda_{23}}\rho_3 = -\frac{1}{2} - \frac{1}{2}\left(\frac{1}{2}\right)(.)^{-.5}(2\lambda_{23} + 2\mu_{21} + 2\lambda_{24} - 2\lambda_{12} - 2\lambda_{14})$$
$$= -\frac{1}{2} - \frac{1}{2}(.)^{-.5}(\lambda_{23} + \mu_{21} + \lambda_{24} - \lambda_{12} - \lambda_{14})$$

$$\frac{\partial}{\partial \lambda_{24}}\rho_3 = -\frac{1}{2} - \frac{1}{2}\left(\frac{1}{2}\right)(.)^{-.5}(2\lambda_{24} + 2\mu_{21} + 2\lambda_{23} - 2\lambda_{12} - 2\lambda_{14})$$
$$= -\frac{1}{2} - \frac{1}{2}(.)^{-.5}(\lambda_{24} + \mu_{21} + \lambda_{23} - \lambda_{12} - \lambda_{14})$$

---

$$\frac{\partial}{\partial \lambda_{12}}\rho_4 = -\frac{1}{2} + \frac{1}{2}\left(\frac{1}{2}\right)(.)^{-.5}(2\lambda_{12} + 2\lambda_{14} + 2\mu_{21} - 2\lambda_{23} - 2\lambda_{24})$$
$$= -\frac{1}{2} + \frac{1}{2}(.)^{-.5}(\lambda_{12} + \lambda_{14} + \mu_{21} - \lambda_{23} - \lambda_{24})$$

$$\frac{\partial}{\partial \lambda_{14}}\rho_4 = -\frac{1}{2} + \frac{1}{2}\left(\frac{1}{2}\right)(.)^{-.5}(2\lambda_{14} + 2\lambda_{12} - 2\mu_{21} - 2\lambda_{23} - 2\lambda_{24})$$
$$= -\frac{1}{2} + \frac{1}{2}(.)^{-.5}(\lambda_{12} + \lambda_{14} - \mu_{21} - \lambda_{23} - \lambda_{24})$$

$$\frac{\partial}{\partial \mu_{21}}\rho_4 = -\frac{1}{2} + \frac{1}{2}\left(\frac{1}{2}\right)(.)^{-.5}(2\mu_{21} + 2\lambda_{12} + 2\lambda_{23} + 2\lambda_{24} - 2\lambda_{14})$$
$$= -\frac{1}{2} + \frac{1}{2}(.)^{-.5}(\mu_{21} + \lambda_{12} + \lambda_{23} + \lambda_{24} - \lambda_{14})$$

$$\frac{\partial}{\partial \lambda_{23}}\rho_4 = -\frac{1}{2} + \frac{1}{2}\left(\frac{1}{2}\right)(.)^{-.5}(2\lambda_{23} + 2\mu_{21} + 2\lambda_{24} - 2\lambda_{12} - 2\lambda_{14})$$
$$= -\frac{1}{2} + \frac{1}{2}(.)^{-.5}(\lambda_{23} + \mu_{21} + \lambda_{24} - \lambda_{12} - \lambda_{14})$$

$$\frac{\partial}{\partial \lambda_{24}}\rho_4 = -\frac{1}{2} + \frac{1}{2}\left(\frac{1}{2}\right)(.)^{-.5}(2\lambda_{24} + 2\mu_{21} + 2\lambda_{23} - 2\lambda_{12} - 2\lambda_{14})$$
$$= -\frac{1}{2} + \frac{1}{2}(.)^{-.5}(\lambda_{24} + \mu_{21} + \lambda_{23} - \lambda_{12} - \lambda_{14})$$

---

$$\frac{\partial}{\partial \theta_h} P_{ij}(t) = t\, e^{\Lambda t}\, d\Lambda$$

$$t\, e^{\Lambda t}\, d\Lambda = t\, e^{\rho_3 t} \begin{bmatrix} \frac{\partial \rho_3}{\partial \lambda_{12}} \\ \frac{\partial \rho_3}{\partial \lambda_{14}} \\ \frac{\partial \rho_3}{\partial \mu_{21}} \\ \frac{\partial \rho_3}{\partial \lambda_{23}} \\ \frac{\partial \rho_3}{\partial \lambda_{24}} \end{bmatrix} + t\, e^{\rho_4 t} \begin{bmatrix} \frac{\partial \rho_4}{\partial \lambda_{12}} \\ \frac{\partial \rho_4}{\partial \lambda_{14}} \\ \frac{\partial \rho_4}{\partial \mu_{21}} \\ \frac{\partial \rho_4}{\partial \lambda_{23}} \\ \frac{\partial \rho_4}{\partial \lambda_{24}} \end{bmatrix}, \text{substitute } t = 1$$

$t\, e^{\Lambda t}\, d\Lambda$ at $t = 1$

$$= e^{\rho_3} \begin{bmatrix} -\frac{1}{2} - \frac{1}{2}(.)^{-.5}(\lambda_{12} + \lambda_{14} + \mu_{21} - \lambda_{23} - \lambda_{24}) \\ -\frac{1}{2} - \frac{1}{2}(.)^{-.5}(\lambda_{14} + \lambda_{12} - \mu_{21} - \lambda_{23} - \lambda_{24}) \\ -\frac{1}{2} - \frac{1}{2}(.)^{-.5}(\mu_{21} + \lambda_{12} + \lambda_{23} + \lambda_{24} - \lambda_{14}) \\ -\frac{1}{2} - \frac{1}{2}(.)^{-.5}(\lambda_{23} + \mu_{21} + \lambda_{24} - \lambda_{12} - \lambda_{14}) \\ -\frac{1}{2} - \frac{1}{2}(.)^{-.5}(\lambda_{24} + \mu_{21} + \lambda_{23} - \lambda_{12} - \lambda_{14}) \end{bmatrix}$$

$$+ e^{\rho_4} \begin{bmatrix} \frac{-1}{2} + \frac{1}{2}(.)^{-.5}(\lambda_{12} + \lambda_{14} + \mu_{21} - \lambda_{23} - \lambda_{24}) \\ \frac{-1}{2} + \frac{1}{2}(.)^{-.5}(\lambda_{12} + \lambda_{14} - \mu_{21} - \lambda_{23} - \lambda_{24}) \\ \frac{-1}{2} + \frac{1}{2}(.)^{-.5}(\mu_{21} + \lambda_{12} + \lambda_{23} + \lambda_{24} - \lambda_{14}) \\ \frac{-1}{2} + \frac{1}{2}(.)^{-.5}(\lambda_{23} + \mu_{21} + \lambda_{24} - \lambda_{12} - \lambda_{14}) \\ \frac{-1}{2} + \frac{1}{2}(.)^{-.5}(\lambda_{24} + \mu_{21} + \lambda_{23} - \lambda_{12} - \lambda_{14}) \end{bmatrix}$$

$$= \begin{bmatrix} -\frac{e^{\rho_3}}{2} - \frac{e^{\rho_4}}{2} + (.)^{-.5}(\lambda_{12} + \lambda_{14} + \mu_{21} - \lambda_{23} - \lambda_{24})\left[\frac{e^{\rho_4}}{2} - \frac{e^{\rho_3}}{2}\right] \\ -\frac{e^{\rho_3}}{2} - \frac{e^{\rho_4}}{2} + (.)^{-.5}(\lambda_{12} + \lambda_{14} - \mu_{21} - \lambda_{23} - \lambda_{24})\left[\frac{e^{\rho_4}}{2} - \frac{e^{\rho_3}}{2}\right] \\ -\frac{e^{\rho_3}}{2} - \frac{e^{\rho_4}}{2} + (.)^{-.5}(\mu_{21} + \lambda_{12} + \lambda_{23} + \lambda_{24} - \lambda_{14})\left[\frac{e^{\rho_4}}{2} - \frac{e^{\rho_3}}{2}\right] \\ -\frac{e^{\rho_3}}{2} - \frac{e^{\rho_4}}{2} + (.)^{-.5}(\lambda_{23} + \mu_{21} + \lambda_{24} - \lambda_{12} - \lambda_{14})\left[\frac{e^{\rho_4}}{2} - \frac{e^{\rho_3}}{2}\right] \\ -\frac{e^{\rho_3}}{2} - \frac{e^{\rho_4}}{2} + (.)^{-.5}(\lambda_{24} + \mu_{21} + \lambda_{23} - \lambda_{12} - \lambda_{14})\left[\frac{e^{\rho_4}}{2} - \frac{e^{\rho_3}}{2}\right] \end{bmatrix}$$

$$v_1 = -\frac{e^{\rho_3}}{2} - \frac{e^{\rho_4}}{2} + (.)^{-.5}(\lambda_{12} + \lambda_{14} + \mu_{21} - \lambda_{23} - \lambda_{24})\left[\frac{e^{\rho_4}}{2} - \frac{e^{\rho_3}}{2}\right]$$
$$v_2 = -\frac{e^{\rho_3}}{2} - \frac{e^{\rho_4}}{2} + (.)^{-.5}(\lambda_{12} + \lambda_{14} - \mu_{21} - \lambda_{23} - \lambda_{24})\left[\frac{e^{\rho_4}}{2} - \frac{e^{\rho_3}}{2}\right]$$
$$v_3 = -\frac{e^{\rho_3}}{2} - \frac{e^{\rho_4}}{2} + (.)^{-.5}(\mu_{21} + \lambda_{12} + \lambda_{23} + \lambda_{24} - \lambda_{14})\left[\frac{e^{\rho_4}}{2} - \frac{e^{\rho_3}}{2}\right]$$
$$v_4 = -\frac{e^{\rho_3}}{2} - \frac{e^{\rho_4}}{2} + (.)^{-.5}(\lambda_{23} + \mu_{21} + \lambda_{24} - \lambda_{12} - \lambda_{14})\left[\frac{e^{\rho_4}}{2} - \frac{e^{\rho_3}}{2}\right]$$
$$v_5 = -\frac{e^{\rho_3}}{2} - \frac{e^{\rho_4}}{2} + (.)^{-.5}(\lambda_{24} + \mu_{21} + \lambda_{23} - \lambda_{12} - \lambda_{14})\left[\frac{e^{\rho_4}}{2} - \frac{e^{\rho_3}}{2}\right]$$

$$M(\theta) = \begin{bmatrix} v_1 \\ v_2 \\ v_3 \\ v_4 \\ v_5 \end{bmatrix} \begin{bmatrix} v_1 & v_2 & v_3 & v_4 & v_5 \end{bmatrix}$$

$$= \begin{bmatrix} v_1^2 & v_1 v_2 & v_1 v_3 & v_1 v_4 & v_1 v_5 \\ v_2 v_1 & v_2^2 & v_2 v_3 & v_2 v_4 & v_2 v_5 \\ v_3 v_1 & v_3 v_2 & v_3^2 & v_3 v_4 & v_3 v_5 \\ v_4 v_1 & v_4 v_2 & v_4 v_3 & v_4^2 & v_4 v_5 \\ v_5 v_1 & v_5 v_2 & v_5 v_3 & v_5 v_4 & v_5^2 \end{bmatrix}$$

According to Klotz and Sharples (1994)

$$\tau = t_r - t_{r-1} = \Delta t \quad \text{and} \quad P_{ij} = \frac{n_{ij}}{n_{i+}}$$

$$S(\theta) = \frac{\partial \log L}{\partial \theta_h} = \sum_{\Delta t \geq 1}^{3} \sum_{i,j=1}^{k} n_{ij} \frac{\partial P_{ij}(\Delta t)/\partial \theta_h}{P_{ij}(\Delta t)}$$
$$= \sum_{\Delta t \geq 1}^{3} \sum_{i,j=1}^{k} n_{ij} \frac{\partial P_{ij}(\Delta t)/\partial \theta_h}{n_{ij}/n_{i+}}$$
$$= \sum_{\Delta t=1}^{3} \sum_{i,j=1}^{k} n_{i+} \frac{\partial P_{ij}(\Delta t)}{\partial \theta_h} = \sum_{\Delta t=1}^{3} \sum_{i,j=1}^{k} n_{i+}\, t\, e^{\Lambda t}\, d\Lambda$$

the resulting vector $t\, e^{\Lambda t}\, d\Lambda$ is scaled by a factor
$$= \left(\frac{n_{ij}(\Delta t)}{P_{ij}(\Delta t)}\right) 8\ \text{times}\ ;\ \text{one for each pdf}$$

the followings are the scalers :

i.e $\frac{n_{11}}{p_{11}} = n_{1+},\ \frac{n_{12}}{p_{12}} = n_{1+},\ \frac{n_{13}}{p_{13}} = n_{1+},\ \frac{n_{14}}{p_{14}} = n_{1+},\ \frac{n_{21}}{p_{21}} = n_{2+},\ \frac{n_{22}}{p_{22}}$
$$= n_{2+},\ \frac{n_{23}}{p_{23}} = n_{2+},\ \frac{n_{24}}{p_{24}} = n_{2+},$$

all at $(\Delta t)$ then the scaled vectors are summed up to get the score function.
This score function is used in $quasi-Newton\ Raphson\ method$:
According to Kalbfliesch and lawless (1985) the second derivative is assumed to be zero, the score function is crossed product with itself and scaled for each pdf with the scalers:
i.e $\frac{n_{1+}}{p_{11}}, \frac{n_{1+}}{p_{12}}, \frac{n_{1+}}{p_{13}}, \frac{n_{1+}}{p_{14}}, \frac{n_{2+}}{p_{21}}, \frac{n_{2+}}{p_{22}}, \frac{n_{2+}}{p_{23}}, \frac{n_{2+}}{p_{24}}$ the scaled matrices are summed up to get the hessian matrix $M(\theta_0)$

$$\frac{\partial^2 Log\ L}{\partial \theta_g \partial \theta_h} = \sum_{\Delta t=1}^{3} \sum_{i,j}^{k} n_{ij} \left[ \frac{\partial^2 P_{ij}(\Delta t)/\partial \theta_g \partial \theta_h}{P_{ij}(\Delta t)} - \frac{\partial P_{ij}(\Delta t)/\partial \theta_g \partial P_{ij}(\Delta t)/\partial \theta_h}{P_{ij}^2(\Delta t)} \right]$$

, where $P_{ij} = \frac{n_{ij}}{n_{i+}}$

$$\frac{\partial^2 Log\ L}{\partial \theta_g \partial \theta_h} = \sum_{\Delta t=1}^{3} \sum_{i,j}^{k} (P_{ij}\ n_{i+}) \left[ \frac{\partial^2 P_{ij}(\Delta t)/\partial \theta_g \partial \theta_h}{P_{ij}(\Delta t)} - \frac{\partial P_{ij}(\Delta t)/\partial \theta_g \partial P_{ij}(\Delta t)/\partial \theta_h}{P_{ij}^2(\Delta t)} \right]$$

$$\frac{\partial^2 Log\ L}{\partial \theta_g \partial \theta_h} = \sum_{\Delta t=1}^{3} \sum_{i,j}^{k} (P_{ij}\ n_{i+}) \left[ \frac{0}{P_{ij}(\Delta t)} - \frac{\partial P_{ij}(\Delta t)/\partial \theta_g \partial P_{ij}(\Delta t)/\partial \theta_h}{P_{ij}^2(\Delta t)} \right]$$

$$\frac{\partial^2 Log\ L}{\partial \theta_g \partial \theta_h} = -\sum_{\Delta t=1}^{3} \sum_{i,j}^{k} n_{i+} \frac{\partial P_{ij}(\Delta t)/\partial \theta_g \partial P_{ij}(\Delta t)/\partial \theta_h}{P_{ij}(\Delta t)}$$

Quasi-Newton Raphson method formula:
$\theta_1 = \theta_0 + M(\theta_0)^{-1} S(\theta_0)$
According to Linda and Klotz (1993); the initial $\theta$ is
$P_{ij} = \frac{n_{ij}}{n_{i+}}\ in\ the\ interval\ \Delta t = 1$
According to Jackson (2019) initial value for a model could be set by supposing that transitions between states take place only at the observation times. If $n_{ij}$ transitions are observed from $state\ i$ to $states\ j$ and a total of $n_i$ transitions from $state\ i$, then $\frac{q_{ij}}{q_{ii}}$ can be estimated by $\frac{n_{ij}}{n_{ii}}$. Then, given a total of $T_i$ years spent in $state\ i$, the mean sojourn time $\frac{1}{q_{ii}}$ can be estimated as $\frac{T_i}{n_i}$. thus, $\frac{n_{ij}}{T_i}$ is a crude estimate of $q_{ij}$.
Substituting in Quasi-Newton method by the initial value, then the score and inverse of the hessian matrix are calculated to give the estimated rates.

$$S(\theta) = (4n_{1+}) \begin{bmatrix} v_1 \\ v_2 \\ v_3 \\ v_4 \\ v_5 \end{bmatrix} + (4n_{2+}) \begin{bmatrix} v_1 \\ v_2 \\ v_3 \\ v_4 \\ v_5 \end{bmatrix}$$

$$= (4n_{1+} + 4n_{2+}) \begin{bmatrix} v_1 \\ v_2 \\ v_3 \\ v_4 \\ v_5 \end{bmatrix}\ ,\ for\ each\ \Delta t$$

$$M(\theta) = (4n_{1+} + 4n_{2+})^2 \begin{bmatrix} v_1 \\ v_2 \\ v_3 \\ v_4 \\ v_5 \end{bmatrix} [v_1\ v_2\ v_3\ v_4\ v_5]$$

$$= (4n_{1+} + 4n_{2+})^2 \begin{bmatrix} v_1^2 & v_1 v_2 & v_1 v_3 & v_1 v_4 & v_1 v_5 \\ v_2 v_1 & v_2^2 & v_2 v_3 & v_2 v_4 & v_2 v_5 \\ v_3 v_1 & v_3 v_2 & v_3^2 & v_3 v_4 & v_3 v_5 \\ v_4 v_1 & v_4 v_2 & v_4 v_3 & v_4^2 & v_4 v_5 \\ v_5 v_1 & v_5 v_2 & v_5 v_3 & v_5 v_4 & v_5^2 \end{bmatrix}$$

$$M(\theta) = (4n_{1+} + 4n_{2+})^2 \begin{bmatrix} v_1^2 & v_1 v_2 & v_1 v_3 & v_1 v_4 & v_1 v_5 \\ v_2 v_1 & v_2^2 & v_2 v_3 & v_2 v_4 & v_2 v_5 \\ v_3 v_1 & v_3 v_2 & v_3^2 & v_3 v_4 & v_3 v_5 \\ v_4 v_1 & v_4 v_2 & v_4 v_3 & v_4^2 & v_4 v_5 \\ v_5 v_1 & v_5 v_2 & v_5 v_3 & v_5 v_4 & v_5^2 \end{bmatrix}$$

$= \begin{bmatrix} O & R \\ X & Y \end{bmatrix}$ then this matrix will be scaled

$M(\theta)$ is a singular matrix

where $O = \begin{bmatrix} v_1^2 & v_1 v_2 & v_1 v_3 \\ v_2 v_1 & v_2^2 & v_2 v_3 \\ v_3 v_1 & v_3 v_2 & v_3^2 \end{bmatrix}$ and it has $O^{-1}$, so $[M(\theta)]^{-1} = \begin{bmatrix} O^{-1} & 0 \\ 0 & 0 \end{bmatrix}$

**II. Mean Sojourn Time**
These times are independent so covariance between them is zero

$$var(s_i) = \left[ \left(q_{ii}(\hat{\theta})\right)^{-2} \right]^2 \sum_{h=1}^{5} \sum_{g=1}^{5} \frac{\partial q_{ii}}{\partial \theta_g} \frac{\partial q_{ii}}{\partial \theta_h} [M(\theta)]^{-1}|_{\theta=\hat{\theta}}$$

$$var(s_i) = \left[ \left(q_{ii}(\hat{\theta})\right)^{-2} \right]^2 \sum_{h=1}^{5} \sum_{g=1}^{5} \left[ \frac{\partial q_{ii}}{\partial \theta_h} \right]^T [M(\theta)]^{-1}|_{\theta=\hat{\theta}} \frac{\partial q_{ii}}{\partial \theta_g}$$

$$\frac{\partial q_{ii}}{\partial \theta_h} = \begin{bmatrix} \frac{\partial q_{11}}{\partial \lambda_{12}} \\ \frac{\partial q_{11}}{\partial \lambda_{14}} \\ \frac{\partial q_{22}}{\partial \mu_{21}} \\ \frac{\partial q_{22}}{\partial \lambda_{23}} \\ \frac{\partial q_{33}}{\partial \lambda_{24}} \end{bmatrix} = \begin{bmatrix} -1 \\ -1 \\ -1 \\ -1 \\ -1 \end{bmatrix} \quad , \quad \frac{\partial q_{ii}}{\partial \theta_g} = \begin{bmatrix} \frac{\partial q_{11}}{\partial \lambda_{12}} \\ \frac{\partial q_{11}}{\partial \lambda_{14}} \\ \frac{\partial q_{22}}{\partial \mu_{21}} \\ \frac{\partial q_{22}}{\partial \lambda_{23}} \\ \frac{\partial q_{33}}{\partial \lambda_{24}} \end{bmatrix} = \begin{bmatrix} -1 \\ -1 \\ -1 \\ -1 \\ -1 \end{bmatrix}$$

$$var(s_i) = \left[ \left(q_{ii}(\hat{\theta})\right)^{-2} \right]^2 \sum_{h=1}^{5} \sum_{g=1}^{5} \left[ \frac{\partial q_{ii}}{\partial \theta_h} \right]^T [M(\theta)]^{-1}|_{\theta=\hat{\theta}} \frac{\partial q_{ii}}{\partial \theta_g}$$

$$var(s_1) = \frac{1}{(\lambda_{12}+\lambda_{14})^4} [-1\ -1\ -1\ -1\ -1][M(\theta)]^{-1}|_{\theta=\hat{\theta}} \begin{bmatrix} -1 \\ -1 \\ -1 \\ -1 \\ -1 \end{bmatrix}$$

$$var(s_2) = \frac{1}{(\mu_{21}+\lambda_{23}+\lambda_{24})^4} [-1\ -1\ -1\ -1\ -1][M(\theta)]^{-1}|_{\theta=\hat{\theta}} \begin{bmatrix} -1 \\ -1 \\ -1 \\ -1 \\ -1 \end{bmatrix}$$

$[M(\theta)]^{-1}|_{\theta=\hat{\theta}}$ as calculated by MLE of rate matrix

**III. State Probability Distribution:**
To get the probability distribution after a certain period of time, the following equation must be solved:
$\pi = \pi(0) P_{ij}(t)$

$$[\pi_1\ \pi_2\ \pi_3\ \pi_4] = [\pi_{1(0)}\ \pi_{2(0)}\ \pi_{3(0)}\ \pi_{4(0)}] \begin{bmatrix} P_{11} & P_{12} & P_{13} & P_{14} \\ P_{21} & P_{22} & P_{23} & P_{24} \\ 0 & 0 & P_{33} & 0 \\ 0 & 0 & 0 & P_{44} \end{bmatrix},$$

where $\pi_{3(0)} = \pi_{4(0)} = 0$, as both are death states
$\pi_1 = \pi_{1(0)} P_{11} + \pi_{2(0)} P_{21}$
$\pi_2 = \pi_{1(0)} P_{12} + \pi_{2(0)} P_{22}$
$\pi_3 = \pi_{1(0)} P_{13} + \pi_{2(0)} P_{23} + \pi_{3(0)} P_{33} = \pi_{1(0)} P_{13} + \pi_{2(0)} P_{23}$
$\pi_4 = \pi_{1(0)} P_{14} + \pi_{2(0)} P_{24} + \pi_{4(0)} P_{44} = \pi_{1(0)} P_{14} + \pi_{2(0)} P_{24}$

**A. Asymptotic Covariance of the Stationary Distribution**
$\frac{\partial}{\partial \theta} F(\theta_h, \pi_i) = \frac{\partial}{\partial \theta_h} (Q' \pi_i) = 0$,

with implicit differentiation,

$\frac{\partial}{\partial \theta_h} F(\theta_h, \pi_i) = \frac{\partial}{\partial \theta_h} (Q' \pi_i) = [Q'] \left[ \frac{\partial}{\partial \theta_h} \pi_i \right] + \pi_i \left[ \frac{\partial}{\partial \theta_h} Q' \right]^T$

, let's call $\pi_i \left[ \frac{\partial}{\partial \theta_h} Q' \right]^T = C(\theta)\ is\ a\ matrix$

$\left[ \frac{\partial}{\partial \theta_h} \pi \right]$ this is a matrix that gives all derivatives of $\pi_1, \pi_2, \pi_3, \pi_4$ with respect to each $\lambda_{12}, \lambda_{14}, \mu_{21}, \lambda_{23}, \lambda_{24}$

$\left[ \frac{\partial}{\partial \theta_h} \pi \right] = -[Q']^{-1} C(\theta)$

$\pi(\theta) = \begin{bmatrix} \pi_1 \\ \pi_2 \\ \pi_3 \\ \pi_4 \end{bmatrix}\ is\ a\ column\ vector$

$C(\theta) = \pi(\theta) \left[ \frac{\partial}{\partial \theta_h} Q' \right]^T = \begin{bmatrix} 0 \\ 0 \\ \pi_3 \\ 1-\pi_3 \end{bmatrix} [1\ 1\ 1\ 1\ 1]$

$= \begin{bmatrix} 0 & 0 & 0 & 0 & 0 \\ 0 & 0 & 0 & 0 & 0 \\ \pi_3 & \pi_3 & \pi_3 & \pi_3 & \pi_3 \\ 1-\pi_3 & 1-\pi_3 & 1-\pi_3 & 1-\pi_3 & 1-\pi_3 \end{bmatrix}$

$$\pi(\theta)\left[\frac{\partial}{\partial \theta_h}Q'\right]^T = C(\theta) \text{ is a matrix of size } (4 \times 5)$$

$$Q' = \begin{bmatrix} -\gamma_1 & \mu_{21} & 0 & 0 \\ \lambda_{12} & -\gamma_2 & 0 & 0 \\ 0 & \lambda_{23} & 0 & 0 \\ \lambda_{14} & \lambda_{24} & 0 & 0 \end{bmatrix}$$

is a singular matrix, and its inverse require calculating the pseudoinverse using SVD
$[Q']^{-1}$ obtained by pseudoinverse is $4$ by $4$ matrix

let: $A(\theta) = \left[\frac{\partial}{\partial \theta_h}\pi_i\right] = -[Q']^{-1}C(\theta)$

$$= \begin{bmatrix} \frac{\partial \pi_1}{\partial \lambda_{12}} & \frac{\partial \pi_1}{\partial \lambda_{14}} & \frac{\partial \pi_1}{\partial \mu_{21}} & \frac{\partial \pi_1}{\partial \lambda_{23}} & \frac{\partial \pi_1}{\partial \lambda_{24}} \\ \frac{\partial \pi_2}{\partial \lambda_{12}} & \frac{\partial \pi_2}{\partial \lambda_{14}} & \frac{\partial \pi_2}{\partial \mu_{21}} & \frac{\partial \pi_2}{\partial \lambda_{23}} & \frac{\partial \pi_2}{\partial \lambda_{24}} \\ \frac{\partial \pi_3}{\partial \lambda_{12}} & \frac{\partial \pi_3}{\partial \lambda_{14}} & \frac{\partial \pi_3}{\partial \mu_{21}} & \frac{\partial \pi_3}{\partial \lambda_{23}} & \frac{\partial \pi_3}{\partial \lambda_{24}} \\ \frac{\partial \pi_4}{\partial \lambda_{12}} & \frac{\partial \pi_4}{\partial \lambda_{14}} & \frac{\partial \pi_4}{\partial \mu_{21}} & \frac{\partial \pi_4}{\partial \lambda_{23}} & \frac{\partial \pi_4}{\partial \lambda_{24}} \end{bmatrix} \text{ is } 4 \text{ by } 5 \text{ matrix}$$

Using multivariate delta method
$var(\pi_i) = A(\theta)var(\theta)A(\theta)'$,
where $var(\theta) = [M(\theta)]^{-1}$ and $i = 1,2,3,4$

**IV. Life Expectancy of Patient in NAFLD Disease Process:**
Solving the following equation to get
$E(\tau_{ik}) = [B]^{-1}Z$
$E(\tau_{ik})$
$$= \begin{bmatrix} \frac{-\gamma_2}{\gamma_1\gamma_2 - \mu_{21}\lambda_{12}} & \frac{-\lambda_{12}}{\gamma_1\gamma_2 - \mu_{21}\lambda_{12}} \\ \frac{-\mu_{21}}{\gamma_1\gamma_2 - \mu_{21}\lambda_{12}} & \frac{-\gamma_1}{\gamma_1\gamma_2 - \mu_{21}\lambda_{12}} \end{bmatrix} \begin{bmatrix} \lambda_{12}\lambda_{23} & \lambda_{12}(\mu_{21}\lambda_{14} + \lambda_{24}\gamma_1) - \lambda_{14}(\mu_{21}\lambda_{12} - \gamma_1\gamma_2) \\ \mu_{21}\lambda_{12} - \gamma_1\gamma_2 & \gamma_1(\mu_{21}\lambda_{12} - \gamma_1\gamma_2) \\ \frac{\gamma_1\lambda_{23}}{\mu_{21}\lambda_{12} - \gamma_1\gamma_2} & \frac{(\mu_{21}\lambda_{14} + \lambda_{24}\gamma_1)}{\mu_{21}\lambda_{12} - \gamma_1\gamma_2} \end{bmatrix}$$

**VI. Hypothetical Numerical Example:**
A study was conducted over 8 years on 310 patients having risk factors to develop NAFLD such as T2DM, obesity and hypertension. The patients were decided to be followed up every 1 year by taking liver biopsy to identify NAFLD cases. The following tables illustrate the counts of transitions in various lengths of time intervals:

Table (1) demonstrates Numbers of observed transitions among states of NAFLD process during different time intervals $\Delta t = 1,2,3$ years

| $\Delta t$ | Transitions among states | | | | | | | |
|---|---|---|---|---|---|---|---|---|
| | (1,1) | (1,2) | (1,3) | (1,4) | (2,1) | (2,2) | (2,3) | (2,4) |
| 1 | 330 | 163 | 45 | 12 | 5 | 185 | 45 | 15 |
| 2 | 70 | 30 | 10 | 1 | 2 | 20 | 13 | 4 |
| 3 | 21 | 8 | 7 | 3 | 1 | 6 | 3 | 1 |
| | 421 | 201 | 62 | 16 | 8 | 211 | 61 | 20 |

Table (2) demonstrates total counts of transitions throughout whole period of the study (8 years)

| | State 1 | State 2 | State 3 | State 4 | Total counts |
|---|---|---|---|---|---|
| State 1 | 421 | 201 | 62 | 16 | 700 |
| State 2 | 8 | 211 | 61 | 20 | 300 |
| State 3 | 0 | 0 | 0 | 0 | 0 |
| State 4 | 0 | 0 | 0 | 0 | 0 |
| Total | 429 | 412 | 123 | 36 | 1000 |

Table (3) demonstrates the observed counts of transitions during time interval $\Delta t = 1$ year

| | State 1 | State 2 | State 3 | State 4 | Total counts |
|---|---|---|---|---|---|
| State 1 | 330 | 163 | 45 | 12 | 550 |
| State 2 | 5 | 185 | 45 | 15 | 250 |
| State 3 | 0 | 0 | 0 | 0 | 0 |
| State 4 | 0 | 0 | 0 | 0 | 0 |
| Total | 335 | 348 | 90 | 27 | 800 |

Table (4) demonstrates the observed counts of transitions during time interval $\Delta t = 2$ years

| | State 1 | State 2 | State 3 | State 4 | Total counts |
|---|---|---|---|---|---|
| State 1 | 70 | 30 | 10 | 1 | 111 |
| State 2 | 2 | 20 | 13 | 4 | 39 |
| State 3 | 0 | 0 | 0 | 0 | 0 |
| State 4 | 0 | 0 | 0 | 0 | 0 |
| Total | 72 | 50 | 23 | 5 | 150 |

Table (5) demonstrates the observed counts of transitions during time interval $\Delta t = 3$ years

| | State 1 | State 2 | State 3 | State 4 | Total counts |
|---|---|---|---|---|---|
| State 1 | 21 | 8 | 7 | 3 | 39 |
| State 2 | 1 | 6 | 3 | 1 | 11 |
| State 3 | 0 | 0 | 0 | 0 | 0 |
| State 4 | 0 | 0 | 0 | 0 | 0 |
| Total | 22 | 14 | 10 | 4 | 50 |

These tables are used to estimate the Q matrix and once the Q matrix is obtained, other statistical indices can be calculated.

**A. Estimating the transition rates:** *(see suppl. Info. In excel sheet)*
Analyzing the rates in first interval $\Delta t = 1$ (first table)
$\rho_3 = -.37443$
$\rho_4 = -.20757$

$$\begin{bmatrix} v_1 \\ v_2 \\ v_3 \\ v_4 \\ v_5 \end{bmatrix} = \begin{bmatrix} -.71195 \\ -.72692 \\ -.5488 \\ -.77332 \\ -.77332 \end{bmatrix}, \rightarrow 4(550) + 4(250) \begin{bmatrix} -.71195 \\ -.72692 \\ -.5488 \\ -.77332 \\ -.77332 \end{bmatrix}$$

$$= \begin{bmatrix} -2278.24439 \\ -2326.14092 \\ -1756.17225 \\ -2474.62015 \\ -2474.62015 \end{bmatrix} = scaled\ S(\theta)$$

$scaled\ S(\theta)[scaled\ S(\theta)]^T = M(\theta)$

$$M(\theta) = \begin{bmatrix} 5190397 & 5299517 & 4000989.6 & 5637789 & 5637789 \\ 5299517 & 5410932 & 4085104 & 5756315 & 5756315 \\ 4000989.6 & 4085104 & 3084141 & 4345859 & 4345859 \\ 5637789 & 5756315 & 4345859 & 6123745 & 6123745 \\ 5637789 & 5756315 & 4345859 & 6123745 & 6123745 \end{bmatrix}$$

$M(\theta)$ is multiplied by $\left(\frac{550^2}{330} + \frac{550^2}{163} + \frac{550^2}{45} + \frac{550^2}{12} + \frac{250^2}{5} + \frac{250^2}{185}\right.$
$\left.+ \frac{250^2}{45} + \frac{250^2}{15}\right) = 53096.45 \cong 53096$

$scaled\ M(\theta)$
$$= \begin{bmatrix} 2.76E+11 & 2.81E+11 & 2.12E+11 & 2.99E+11 & 2.99E+11 \\ 2.81E+11 & 2.87E+11 & 2.17E+11 & 3.06E+11 & 3.06E+11 \\ 2.12E+11 & 2.17E+11 & 1.64E+11 & 2.31E+11 & 2.31E+11 \\ 2.99E+11 & 3.06E+11 & 2.31E+11 & 3.25E+11 & 3.25E+11 \\ 2.99E+11 & 3.06E+11 & 2.31E+11 & 3.25E+11 & 3.25E+11 \end{bmatrix}$$

$[scaled\ M(\theta)]^{-1}$
$$= \begin{bmatrix} 1.34E-09 & -1.6E-09 & 3.78E-10 & 0 & 0 \\ 6.14E-10 & -3.1E-09 & 3.27E-09 & 0 & 0 \\ -2.6E-09 & 6.14E-09 & -4.82E-09 & 0 & 0 \\ 0 & 0 & 0 & 0 & 0 \\ 0 & 0 & 0 & 0 & 0 \end{bmatrix}$$

$[scaled\ M(\theta)]^{-1} \times scaled\ S(\theta) = [scaled\ M(\theta)]^{-1} \times scaled\ S(\theta)$

$$\theta_1 = \theta_0 + [scaled\ M(\theta)]^{-1} scaled\ S(\theta) = \begin{bmatrix} .30000001 \\ .02200007 \\ .0200001 \\ .18 \\ .06 \end{bmatrix}$$

It is observed that this rate vector is almost the initial rate vector. No need for second iteration, because the difference between $\theta_1$ and $\theta_0$ is zero as shown from Quasi-Newton equation. Repeating this procedure for $\Delta t = 2$ and $\Delta t = 3$ will give the following vectors respectively (substitute for t=2 and t=3 in their intervals):

for $\Delta t = 2$
$$\rightarrow \vec{\theta} = \begin{bmatrix} .27 \\ .009 \\ .05 \\ .333 \\ .103 \end{bmatrix} \text{ is obtained in the first iteration}$$

with a difference from the initial values
$$= \begin{bmatrix} 1.42E-09 \\ -7.1E-08 \\ -9.6E-08 \\ 0 \\ 0 \end{bmatrix}$$

while for $\Delta t = 3$
$$\rightarrow \vec{\theta} = \begin{bmatrix} .206172 \\ .077985 \\ .091339 \\ .273 \\ .091 \end{bmatrix} \text{ obtained in second iteration}$$

with a difference from the initial values

$$= \begin{bmatrix} 1.17E-03 \\ 9.87E-04 \\ 3.40E-04 \\ 0 \\ 0 \end{bmatrix}$$

As noted from this procedure in all time intervals, the initial values are almost the estimated values regardless of the interval.
If the scaled score function in each iteration is weighted according to the contribution of the counts of transitions in this interval to the whole number of transitions (1000 transitions) and summed up, this will give

$$(.8)\begin{bmatrix}.3\\.022\\.02\\.18\\.06\end{bmatrix} + (.15)\begin{bmatrix}.3\\.022\\.02\\.18\\.06\end{bmatrix} + (.05)\begin{bmatrix}.206\\.078\\.091\\.273\\.091\end{bmatrix} = \begin{bmatrix}.2908\\.02285\\.02805\\.2076\\.068\end{bmatrix}$$

Also the weighted sum of the inversed scaled hessian matrix should be used as the variance-covariance matrix of parameter $\theta$

$$[scaled\ M(\theta)]^{-1} = \begin{bmatrix}.061475 & -.04645 & -.01585 & 0 & 0\\ -.04645 & .037836 & -.00613 & 0 & 0\\ -.01585 & -.00613 & .123658 & 0 & 0\\ 0 & 0 & 0 & 0 & 0\\ 0 & 0 & 0 & 0 & 0\end{bmatrix}$$

**B. Calculating the Mean Sojourn Time:**
It is the average amount of time spent by a patient in the state:

$$E(s_1) = \frac{1}{\lambda_{12}+\lambda_{14}} = \frac{1}{.2908+.02285} = 3.19\ year$$

$$E(s_2) = \frac{1}{\mu_{21}+\lambda_{23}+\lambda_{24}} = \frac{1}{.02805+.2076+.068} = 3.29\ year$$

**C. Calculating the Variance of Sojourn Time:**

$$var(s_i) = \left[\left(q_{ii}(\hat{\theta})\right)^{-2}\right]^2 \sum_{h=1}^{5}\sum_{g=1}^{5}\left[\frac{\partial q_{ii}}{\partial \theta_h}\right]^T [M(\theta)]^{-1}|_{\theta=\hat{\theta}} \frac{\partial q_{ii}}{\partial \theta_g}$$

where $[M(\theta)]^{-1}$ is the weighted sum of inverted scaled hessian matrix

$$var(s_1) = \frac{1}{(.2908+.02285)^4}\begin{bmatrix}-1 & -1 & -1 & -1 & -1\end{bmatrix}[M(\theta)]^{-1}|_{\theta=\hat{\theta}}\begin{bmatrix}-1\\-1\\-1\\-1\\-1\end{bmatrix}$$

$$= 8.898$$

$$var(s_2)$$

$$= \frac{1}{(.02805+.2076+.068)^4}\begin{bmatrix}-1 & -1 & -1 & -1 & -1\end{bmatrix}[M(\theta)]^{-1}|_{\theta=\hat{\theta}}\begin{bmatrix}-1\\-1\\-1\\-1\\-1\end{bmatrix}$$

$$= 10.129$$

**D. State Probability Distribution:**
Once the rate matrix is obtained, these estimated rates are substituted into the calculated Pdf's from the solved differential equations to get the state probability distribution at any point in time as well as the expected number of patients.
Studying a cohort of 3000 patients with the initial distribution $[.7\ .3\ 0\ 0]$, and initial numbers of patients in each state are $[2100\ 900\ 0\ 0]$.
At 1 year the state probability distribution is approximately:

$$P(1) = [.7\ .3\ 0\ 0]\begin{bmatrix}.734 & .214 & .025 & .027\\.021 & .741 & .179 & .059\\0 & 0 & 1 & 0\\0 & 0 & 0 & 1\end{bmatrix}$$
$$= [.52\ .372\ .071\ .037]$$

And the expected numbers of patients in each state is:

$$[2100\ 900\ 0\ 0]\begin{bmatrix}.734 & .214 & .025 & .027\\.021 & .741 & .179 & .059\\0 & 0 & 1 & 0\\0 & 0 & 0 & 1\end{bmatrix}$$
$$= [1559\ 1117\ 214\ 110]$$

At 20 years the state probability distribution is approximately:

$$P(20) = [.7\ .3\ 0\ 0]\begin{bmatrix}.006 & .019 & .675 & .3\\.002 & .006 & .742 & .25\\0 & 0 & 1 & 0\\0 & 0 & 0 & 1\end{bmatrix}$$
$$= [.0048\ .0151\ .6951\ .285]$$

And the expected numbers of patients in each state is:

$$[2100\ 900\ 0\ 0]\begin{bmatrix}.006 & .019 & .675 & .3\\.002 & .006 & .742 & .25\\0 & 0 & 1 & 0\\0 & 0 & 0 & 1\end{bmatrix}$$
$$= [15\ 45\ 2085\ 855]$$

At 60 years the state probability distribution is approximately:

$$P(60) = [.7\ .3\ 0\ 0]\begin{bmatrix}0 & 0 & .7 & .3\\0 & 0 & .75 & .25\\0 & 0 & 1 & 0\\0 & 0 & 0 & 1\end{bmatrix} = [0\ 0\ .715\ .285]$$

And the expected numbers of patients in each state is:

$$[2100\ 900\ 0\ 0]\begin{bmatrix}0 & 0 & .7 & .3\\0 & 0 & .75 & .25\\0 & 0 & 1 & 0\\0 & 0 & 0 & 1\end{bmatrix} = [0\ 0\ 2145\ 855]$$

**E. Asymptotic Covariance of the Stationary Distribution :**
At 60 years the state probability distribution is $[0\ 0\ .715\ .285]$, so to calculate the
$\left[\frac{\partial}{\partial \theta_h}\pi\right]$ , $C(\theta)$ matrix is calculated as in the following steps:

$$C(\theta) = \pi(\theta)\left[\frac{\partial}{\partial \theta_h}Q'\right]^T = \begin{bmatrix}0\\0\\.7\\.3\end{bmatrix}[1\ 1\ 1\ 1\ 1]$$

$$= \begin{bmatrix}0 & 0 & 0 & 0 & 0\\0 & 0 & 0 & 0 & 0\\.7 & .7 & .7 & .7 & .7\\.3 & .3 & .3 & .3 & .3\end{bmatrix}$$

then $\left[\frac{\partial}{\partial \theta_h}\pi\right] = -[Q']^{-1}C(\theta)$, $-[Q']^{-1}$ is calculated taking into account that $Q'$ is a singular matrix and its inverse (the pseudoinverse) is obtained via singular value decomposition (SVD).

$$Q' = \begin{bmatrix}-.31365 & .02805 & 0 & 0\\.2908 & -.30365 & 0 & 0\\0 & .2076 & 0 & 0\\.02285 & .068 & 0 & 0\end{bmatrix}, \text{by SVD}$$

$$[Q']^+ = \begin{bmatrix}-2.48429 & .71355 & 1.1887 & .58205\\-1.48753 & -1.6734 & 2.2825 & .87848\\0 & 0 & 0 & 0\\0 & 0 & 0 & 0\end{bmatrix}$$

$$A(\theta) = -[Q']^+ C(\theta)$$
$$= \begin{bmatrix}-1.01422 & -1.01422 & -1.01422 & -1.01422 & -1.01422\\-1.8779 & -1.8779 & -1.8779 & -1.8779 & -1.8779\\0 & 0 & 0 & 0 & 0\\0 & 0 & 0 & 0 & 0\end{bmatrix}$$

$$A(\theta)[scaled\ M(\theta)]^{-1}[A(\theta)]^T = \begin{bmatrix}.088589 & .16401 & 0 & 0\\.16401 & .30368 & 0 & 0\\0 & 0 & 0 & 0\\0 & 0 & 0 & 0\end{bmatrix}$$

**F. Life Expectancy of the Patient (mean time to absorption):**
$E(\tau_{ik}) = [B]^{-1}Z$
$$E(\tau_{ik}) = \begin{bmatrix}-3.48691 & -3.33935\\-.32212 & -3.60174\end{bmatrix}\begin{bmatrix}-.69325 & -.30675\\-.74772 & -.25228\end{bmatrix}$$
$$= \begin{bmatrix}4.9142 & 1.9121\\2.9164 & 1.0074\end{bmatrix}$$

$E(\tau_{13}) = 4.9142\ years$
$E(\tau_{14}) = 1.9121\ years$
$E(\tau_{23}) = 2.9164\ years$
$E(\tau_{24}) = 1.0074\ years$